\documentclass[12pt]{article}
\usepackage[papersize={8.5in,11in},textwidth=6.5in,bottom=1in]{geometry}

\usepackage{sectsty}
\sectionfont{\normalsize\bfseries}
\subsectionfont{\normalsize\sf}
\usepackage[mathcal]{euscript}

\usepackage{bm}                     
\usepackage{amsfonts}
\usepackage{amssymb}
\usepackage{amsmath}
\usepackage{amsthm}
\usepackage{graphicx}               
\usepackage{natbib}                 
\usepackage{colortbl}               
\usepackage{booktabs}
\usepackage{multicol} 
\usepackage{amssymb,amsmath}

\usepackage[utf8]{inputenc}
\usepackage{graphicx,color}
\newtheorem{theorem}{Theorem}

\newtheorem{proposition}{Proposition}

\bibliography{referencias}

\newcommand{\innerk}[2]{\langle #1 , #2  \rangle_K }
\newcommand{\innerl}[2]{\langle #1 , #2  \rangle_{L^2} }

\newcommand{\normk}[1]{\parallel #1 \parallel_K}

\usepackage{setspace}
\definecolor{violet}{rgb}{.7,.4,1}
\def\spacingset#1{\renewcommand{\baselinestretch}%
	{#1}\small\normalsize} \spacingset{1}

\begin{document}

\begin{center}
\large \bf  On the use of reproducing kernel Hilbert spaces in functional classification\normalsize
\end{center}
\normalsize

\begin{center}
  Jos\'e R. Berrendero, Antonio Cuevas, Jos\'e L. Torrecilla \\
  Departamento de Matem\'aticas\\
  Universidad Aut\'onoma de Madrid, Spain
\end{center}

\begin{abstract}

\footnotesize The H\'ajek-Feldman dichotomy establishes that two Gaussian measures are either mutually absolutely continuous with respect to each other (and hence there is a Radon-Nikodym density for each measure with respect to the other one) or mutually singular. Unlike the case of finite dimensional Gaussian measures, there are non-trivial examples of both situations when dealing with Gaussian stochastic processes.  This paper provides:

(a) Explicit expressions for the optimal (Bayes) rule and the minimal classification error probability in several relevant problems of supervised binary classification of mutually absolutely continuous Gaussian processes. The approach relies on some classical results in the theory of Reproducing Kernel Hilbert Spaces (RKHS).

(b) An interpretation, in terms of mutual singularity,  for the ``near perfect classification'' phenomenon described by \cite{del12}.  We show that the asymptotically optimal rule proposed by these authors can be identified with the sequence of optimal rules for an approximating sequence of classification problems  in the absolutely continuous case. 

(c) A new model-based method for variable selection in binary classification problems, which arises in a very natural way from the explicit knowledge of the RN-derivatives and the underlying RKHS structure. Different classifiers might be used from the selected variables. In particular, the classical, linear finite-dimensional Fisher rule turns out to be consistent under some standard conditions on the underlying functional model.  

\end{abstract}

\noindent {\bf Keywords:} absolutely continuous processes, Radon--Nikodym derivatives, singular processes,  supervised functional classification, variable selection.

\noindent {\bf AMS 2010 subject classifications:} Primary 62H30; secondary 62G99.

\doublespacing
\section{Introduction}\label{sec:intro}

In the booming field of statistics with functional data [see \cite{cue14} for a recent survey], the computational and numerical aspects, as well as the real data applications, have had (understandably) a major role so far. However, the underlying probabilistic theory, connecting the models which generate the data (i.e., the stochastic processes) with the statistical functional methods is far less developed. 
The present work is an attempt to contribute to that connection. Our conclusions will present both theoretical and practical aspects. Roughly speaking, our aim is to prove that in the field of supervised functional classification, there are many useful underlying models (defined in terms of appropriate stochastic processes) for which the expression of the optimal rule can be explicitly given. This will also lead to a natural procedure for variable selection in these models. We are also able to shed some light on the interesting phenomenon of ``near perfect classification'', discussed by \cite{del12}. This phenomenon does not appear (except for trivial or artificial cases) in the classical finite-dimensional classification theory.

\subsection{The framework: supervised classification and absolute continuity}\label{scp} 

We are concerned here with the problem of binary functional supervised classification. Throughout the paper $X=X(t)=X_t=X(t,\omega)$ will denote a stochastic process with $t\in I$, for some compact interval $I$. Unless otherwise specified we will assume $I=[0,T]$, with $T>0$.  This process can be observed in two populations identified by the random ``label'' variable $Y$; the conditional distributions of $X|Y=i$  for $i=0,1$, denoted by $P_i$,
are assumed to be Gaussian. 

As usual in the supervised classification setting, the aim is to classify an ``unlabelled'' observation $X$ according to whether it comes from $P_0$ or from $P_1$. A classification rule is just a measurable function $g:{\mathcal X}\rightarrow \{0,1\}$, where ${\mathcal X}$ is the space of trajectories of the process $X$.

The expression $P_1<<P_0$ indicates that $P_1$ is absolutely continuous with respect to $P_0$ (i.e. $P_0(A)=0$ entails
$P_1(A)=0$). Note that, from the H\'ajek-Feldman dichotomy for Gaussian measures  \citep{fel58}, $P_1<<P_0$  implies also $P_0<<P_1$, so that both measures are in fact mutually absolutely continuous (or ``equivalent''). This is often denoted $P_1\sim P_0$.

When $P_0$ and $P_1$ are completely known in advance and $P_1<<P_0$,  it can be shown that the optimal classification rule 
(often called \it Bayes rule\/\rm)  is
\begin{equation}
g^*(x)={\mathbb I}_{\left\{\eta(x)>1/2\right\}}={\mathbb I}_{\left\{\frac{dP_1(x)}{dP_0}>\frac{1-p}{p}\right\}},\label{bay}
\end{equation}
where ${\mathbb I}$ denotes the indicator function, $\eta(x)={\mathbb P}(Y=1|X=x)={\mathbb E}(Y|X=x)$,  $p={\mathbb P}(Y=1)$ and $\frac{dP_1(x)}{dP_0}$ is the Radon-Nikodym derivative of $P_1$ with respect to $P_0$. The corresponding minimal ``classification error'' (i.e., the misclassification probability) $L^*={\mathbb P}(g^*(X)\neq Y)$ is called \it Bayes error\rm; see, e.g., \cite{dev96} for general background and \cite{bai11a} for additional details on the functional case.

If the Radon-Nikodym derivative $\frac{dP_1(x)}{dP_0}$ is explicitly known, there is not much else to be said. However, in practice, this is not usually the case. Even if the general expression of $\frac{dP_1(x)}{dP_0}$
is known, it typically depends on the covariance $K(s,t)=\mbox{Cov}(X(s),X(t))$ and mean functions $m_i(t)={\mathbb E}(X(t)|Y=i)$.  

The term ``supervised'' accounts for the fact that, in any case,  a data set of ``well-classified'' independent observations ${\mathcal D}_n=((X_1,Y_1),\ldots,(X_n,Y_n))$ from $(X,Y)$ is assumed to be available beforehand. So, the classification rules are in fact constructed in terms of the sample data ${\mathcal D}_n$. Throughout the paper, the functional data $X=X(t)$ are supposed to be ``densely observed''; see, e.g., \citet[Sec 2.1]{cue14}. A common strategy is to use these data to estimate the optimal rule (\ref{bay}). This is the so-called \it plug--in approach\rm. It is often implemented in a non-parametric way (e.g., estimating $\eta(x)$ by a nearest-neighbour estimator) which does not require much information on the precise structure of $\eta(x)$ or $\frac{dP_1(x)}{dP_0}$. However, in some other cases we have a quite precise information on the structure of  $\frac{dP_1(x)}{dP_0}$, so that we can take advantage of this information to get better plug-in estimators of $g^*(x)$. .
\subsection{Some especial characteristics of classification with functional data. The aims of this work}\label{especial}

It can be seen from the above paragraphs that the supervised classification problem can be stated, with almost no formal difference, either in the ordinary finite-dimensional situation (where $X$ takes values on the Euclidean space ${\mathcal X}={\mathbb R}^d$) or in the functional case (where $X$ is a stochastic process). In spite of these formal analogies, the passage to an infinite-dimensional (functional) sample space ${\mathcal X}$ entails some very important challenges. For example, the classical Fisher linear rule,  which is still very popular in the finite-dimensional setting, cannot be easily adapted to the functional case (see, \cite{bai11b} for more details and references). However, we are more concerned here with another crucial difference, namely the lack of a natural ``dominant'' measure in functional spaces, playing a similar role to that of Lebesgue measure in ${\mathbb R}^d$. If we are working with Gaussian measures in ${\mathbb R}^d$, the optimal rule (\ref{bay}) can be established (using the chain rule for Radon-Nikodym derivatives) in terms of the ordinary (Lebesgue) densities of $P_0$ and $P_1$. In the functional case, we are forced to work with the ``mutual'' Radon-Nikodym derivatives $dP_1/dP_0$, provided that $P_1<<P_0$. Usually these derivatives are not  easy to calculate or to work with. However,  in some important examples they are explicitly known and reasonably easy to handle. 

So first, we give and interpret explicit expressions for the optimal (Bayes) classification rule in some relevant cases with $P_1<<P_0$. Similar ideas are developed in \cite{bai11a}
and \cite{cad13} but, unlike these references, our approach here relies heavily on the theory of Reproducing Kernel Hilbert Spaces (RKHS). See Sections \ref{back} and \ref{abs} below. 

In the second place, we consider  the mutually singular case $P_1\bot P_0$, i.e., when there exists a Borel set $A$ such that $P_0(A)=1$ and $P_1(A)=0$. Note that  this mutually singular (or ``orthogonal'') case is rarely found in the finite-dimensional classification setting, except in a few trivial or artificial cases. However, in the functional setting (that is, when $P_1$ and $P_0$ are distributions of stochastic processes) the singular case is an important, very common situation. As we argue in Section \ref{bot}, this mutual singularity notion is behind the near perfect classification phenomenon described in \cite{del12}; see also \cite{cues16}. The point is to look at this phenomenon from a slightly different (coordinate free) RKHS perspective. We also show that an approximately optimal (``near perfect'') classification rule to discriminate between $P_0$ and $P_1$ when $P_1\perp P_0$, can be obtained in terms of the optimal rules  of a sequence of problems $(P_0^n,P_1^n)$ with 
$P_1^n<<P_0^n$. 

Third, in Section \ref{sec:variableselection} we propose an RKHS-based variable selection mechanism (RK-VS hereafter). Unlike other popular variable selection methods in classification (see, e.g., \cite{ber15a}) this new proposal allows the user to incorporate, in a flexible way, different amounts of information (or assumptions) on the underlying model. We also provide a closely related  linear  classifier denoted henceforth by RK-C.
As shown in Section \ref{sec:experiments}, both the variable selection method  and the associated classifier perform very well and are clearly competitive compared to several natural alternatives.  We also argue, as an important additional advantage, the simplicity and ease of interpretation of the RKHS-based procedures.

All proofs and some details about de simulation models are given in the \textit{Supplementary material} document. 

\section{Radon-Nikodym densities for Gaussian processes: some background}\label{back}

In the following paragraphs we review, for posterior use, some results regarding the explicit calculation of Radon-Nikodym  derivatives of Gaussian processes in the convenient setting provided by the theory of Reproducing Kernel Hilbert Spaces.

\subsection{RKHS}\label{sub:rkhs}

We first need to recall some very basic facts on the RKHS theory; see \cite{ber04}, \citet[Appendix F]{jan97} for background. 

Given a symmetric positive-semidefinite function $K(s,t)$, defined on $[0,T]\times [0,T]$ (in our case $K$ will be the covariance function of a process), let us define the space ${\mathcal H}_0(K)$ of all real functions which can be expressed as finite linear combinations of type $\sum_ia_iK(\cdot,t_i)$ (i.e., the linear span of all functions $K(\cdot, t)$). In 
${\mathcal H}_0(K)$ we consider the inner product 
$\langle f,g\rangle_K=\sum_{i,j}\alpha_i\beta_jK(s_j,t_i)$, 
where $f(x)=\sum_i\alpha_i K(x,t_i)$ and $g(x)=\sum_j\beta_jK(x,s_j)$.

Then, the RKHS associated with $K$, ${\mathcal H}(K)$,  is defined as the completion of 
${\mathcal H}_0(K)$. More precisely, ${\mathcal H}(K)$ is the set of functions $f:[0,T]\rightarrow {\mathbb R}$ which can be obtained as $t$ pointwise limit of a Cauchy sequence $\{f_n\}$ of functions in ${\mathcal H}_0(K)$. The theoretical motivation for this definition is the well-known Moore-Aronszajn Theorem (see \cite{ber04}, p. 19).
The functions in ${\mathcal H}(K)$ have the ``reproducing property'' 
$f(t)=\langle f,K(\cdot,t)\rangle_K$.

If $\{X_t, t\in[0,T]\}$ is an $L^2$-process (i.e. ${\mathbb E}(X_t^2)<\infty$, for all $t$) with covariance function $K(s,t)$, the natural Hilbert space
associated with this process, $\bar {\mathcal L}(X)$ is the closure (in $L^2$) of the linear span ${\mathcal L}(X)={\mathcal L}(X_t,\ t\in[0,T])$. The so-called \it Lo\`eve Representation Theorem\/ \rm \citep[p. 65]{ber04} establishes that the spaces ${\mathcal L}(X)$ and ${\mathcal H}(K)$ are \it congruent\rm. More precisely, the natural transformation $\Psi(\sum_ia_iX_{t_i})=\sum_i a_i K(\cdot,t_i)$ defines in fact, when extended by continuity, a congruence (that is an isomorphism which preserves the inner product) between $\bar{\mathcal L}(X)$ and ${\mathcal H}(K)$.  Two interesting consequences of Lo\`eve's result are: first, if a linear map $\phi$, from $\bar{\mathcal L}(X)$ to ${\mathcal H}(K)$, fulfils ${\mathbb E}(\phi^{-1}(h)X_t)=h(t)$, for all $h\in {\mathcal H}(K)$, then  $\phi$ coincides with the congruence $\Psi$ which maps $X_t$ to $K(t,\cdot)$. Second, ${\mathcal H}(K)$ coincides with the space of functions of the form  $h(t)={\mathbb E}(X_tU)$, for some $U\in\bar{\mathcal L}(X)$.

Thus, in a very precise way,  ${\mathcal H}(K)$ can be seen as the ``natural Hilbert space'' associated with a process $\{X(t), t\in[0,T]\}$. In fact, as we will next see, the space ${\mathcal H}(K)$ is deeply involved in some relevant probabilistic and statistical notions.

\subsection{RKHS and Radon-Nikodym derivatives. Parzen's Theorem}\label{sub:parzen}

The following result is a slightly simplified version of Theorem 7A in \cite{par61}; see also \cite{par62}. It will be particularly useful in the rest of this paper. 

\begin{theorem}
	\label{theorem:parzen} \citep[Th. 7A]{par61}. Let us denote by $P_1$ the distribution of a Gaussian process $\{X(t),\ t\in[0,T]\}$, with continuous trajectories,  mean function  denoted by $m=m(t)={\mathbb E}(X(t))$ and continuous covariance function denoted by $K(s,t)=\mbox{Cov}(X(s),X(t))$. Let $P_0$ be  the distribution of another Gaussian process with the same covariance function and with mean function identically 0. Then, $P_1<<P_0$ if and only if the mean function $m$ belongs to the space ${\mathcal H}(K)$. In this case,
	\begin{equation}
	\frac{dP_1(X)}{dP_0}=\exp\left(\langle X,m\rangle_K-\frac{1}{2}\langle m,m\rangle_K \right).\label{parzendens}
	\end{equation}
	In the case $m\notin  {\mathcal H}(K)$, we have $P_1\bot P_0$. 
	
\end{theorem}

\noindent \it Some remarks on this result\rm.

	(a) Note that, except for trivial cases, the trajectories $x$ of the process $X(t)$ are not included, with probability one, in ${\mathcal H}(K)$; see, e.g., \cite[p. 66]{ber04} and \cite{luk01} for details. Thus, the expression $\langle X,m\rangle_K$ is defined a.s. as the random variable $\Psi^{-1}(m)$, where $\Psi^{-1}$ is the inverse of the above defined congruence $\Psi:\bar{\mathcal L}(X)\to {\mathcal H}(K)$ which maps $X_t$ to $K(t,\cdot)$. This definition of $\langle X,m\rangle_K$ in terms of a congruence, is strongly reminiscent of the definition of the It\^o's stochastic integral. 
	
	(b) As a matter of fact, $\langle X,m\rangle_K$ can be seen as a stochastic integral. To see this consider the classical case where $X(t)=B(t)$ is the standard Brownian Motion, $K(s,t)=\min(s,t)$. Then, it can be seen that ${\mathcal H}(K)$ coincides with the so-called Dirichlet space ${\mathcal D}[0,T]$ of those real functions $g$ on $[0,T]$ such that there exists $g^\prime$ almost everywhere in $[0,T]$ with
	$g^\prime\in L^2[0,T]$, and 
	$g(t)=\int_0^tg^\prime(s)ds$.
	The norm in ${\mathcal D}[0,T]$ is defined by $\Vert g\Vert_K=\left(\int_0^T g'^2(t)dt\right)^{1/2}$.
	Likewise, the inverse congruence  $\langle X,m\rangle_K$ can also be expressed as the stochastic integral 
	$\int_0^Tm'(s)dB(s)$.
	
	Thus, Theorem \ref{theorem:parzen} 
	can be seen as an extension of the classical Cameron-Martin Theorem \cite[p. 24]{mor10}, which is stated  for $X(t)=B(t)$. It also coincides with \citet[Th. 1]{she66}, when applied to the homoscedastic case in which $P_0$ and $P_1$ are the distributions of $X(t)$ and $m(t)+X(t)$, respectively.
	
	(c) Some additional references on  Radon-Nikodym derivatives in function spaces are \cite{var61,var64}, \cite{kai71} and \cite{seg75}, among others.

\section{Classification of absolutely continuous Gaussian processes}\label{abs} 
In this section we consider the supervised classification problem, as stated in Subsection \ref{scp}, under the following general model 
\begin{equation}
\label{modelogeneral}
\left\{ \begin{array}{ll}
P_0: & m_0(t) + \epsilon_0(t) \\
P_1: & m_1(t)+\epsilon_1(t)
\end{array}  \right. \hspace{10pt}, 
\end{equation}
where, for $i=0,1$, $\{\epsilon_i(t),\ t\in I\}$ are ``noise processes'' with mean 0 and continuous trajectories, and $m_i(t)$,
$t\in I$ are some  continuous functions 
defining the respective ``trends'' of $P_0$ and $P_1$. We will take $I=[0,T]$ unless otherwise stated.

The following result provides the expression of the Bayes (optimal) rule and the corresponding minimal error probability for this case, under the usual assumption of homoscedasticity. While the proof is a simple consequence of Theorem \ref{theorem:parzen} and Theorem 1 in \cite{bai11a}, this result will be essential in the rest of the paper. 

\begin{theorem}\label{theorem:optParzen}
	In the classification problem under the model (\ref{modelogeneral}) assume 
	\begin{itemize}
		\item[(a)] the noise processes $\epsilon_i$ are both Gaussian with continuous trajectories and common continuous covariance function $K(s,t)$. 
		\item[(b)] $m:=m_1-m_0\in {\mathcal H}(K)$, where ${\mathcal H}(K)$ denotes the RKHS associated with $K$.
	\end{itemize}
	Then, the optimal Bayes rule is given by $g^*(X)=\mathbb{I}_{\{\eta^*(X)>0\}}$, where
	{\small \begin{equation}
		\label{eq:BayesParzen}
		\eta^*(X) = 
		\innerk{X-m_0}{m} - \frac{1}{2}\normk{m}^2 - \log\left(\frac{1-p}{p} \right),
		\end{equation}}
	and $\Vert \cdot\Vert_K$ denotes the norm in the space ${\mathcal H}(K)$. 
	
	Also, the corresponding optimal classification error $L^*={\mathbb P}(g^*(X)\neq Y)$ 
	is
	$$L^* = (1-p) \Phi\left(-\frac{\normk{m}}{2}-\frac{1}{\normk{m}}\log\left(\frac{1-p}{p} \right) \right) + p \Phi\left(-\frac{\normk{m}}{2}+\frac{1}{\normk{m}}\log\left(\frac{1-p}{p} \right) \right),
	$$
	where $\Phi$ is the cumulative distribution function of a standard normal random variable. When $p=1/2$, we have 
	$L^* = 1-\Phi\left(\frac{\normk{m}}{2} \right)$.

\end{theorem}

If we compare this result with he optimal rule given for a similar problem in Theorem 1 of the paper \cite{del12}, we see that (\ref{eq:BayesParzen}) does not explicitly depends on the eigenvalues and eigenvectors of the covariance operator. As a counterpart, the general expression expression (\ref{eq:BayesParzen}) is given in terms of the ``stochastic integral'' $\langle X,m\rangle_K$. We will comment on this in more detail in the next section.

%
%
%
%
%
%
%
%

\section{Classification of Gaussian processes: another look at the ``near perfect classification'' phenomenon} \label{bot}

The starting point in this section is again the classification problem between the Gaussian processes $P_0$ and $P_1$ defined in
(\ref{modelogeneral}), where $\epsilon_0$ and $\epsilon_1$ are identically distributed according to  the Gaussian process $\epsilon(t)$ with covariance function $K(s,t)=\mathbb{E}(\epsilon(s)\epsilon(t))$. The mean functions are $m_0(t)=0$ and $m_1(t)=\sum_{j=1}^\infty \mu_j \phi_j(t)$, where the $\phi_j$ are the eigenfunctions of the Karhunen-Loève  expansion of $K$, that is
$K(s,t)=\sum_{j=1}^\infty \theta_j \phi_j(s)\phi_j(t).$

Let us assume for simplicity that the prior probability is ${\mathbb P}(Y=1)=1/2$. 
This model has been considered by \cite{del12}. In short, these authors provide the explicit expression of the optimal rule under the assumption $\sum _{j=1}^\infty \theta_j^{-2}\mu_j^2<\infty$. In addition, they find that, when $\sum _{j=1}^\infty \theta_j^{-1}\mu_j^2=\infty$, the classification is ``near perfect'' in the sense that one may construct a rule with an arbitrarily small classification error. To be more specific, the classification rule they propose is the so-called
``centroid classifier'', $T_n$, defined by
$T_n(X)=1$ if and only if $D^2(X,\bar{X}_1)-D^2(X,\bar{X}_0)<0$,
where $\bar{X}_0,\,\bar{X}_1$ denote the sample means of the training data from $P_0$ and $P_1$ and 
$D(X,\bar{X}_j)=|\innerl{X}{\psi} - \innerl{\bar{X}_j}{\psi}|$, with $\innerl{X}{\psi}=\int_0^T X(t)\psi(t)dt$
and $\psi(t)=\sum _{j=1}^\infty \theta_j^{-1}\mu_j\phi_j(t)$. Of course, this requires $\psi\in L^2$
which (from Parseval's identity) amounts to $\sum _{j=1}^\infty \theta_j^{-2}\mu_j^2<\infty$.
 Then, the asymptotic version of the classifier $T_n$
under the assumed model  is
\begin{equation}
T^0(X)=1, \mbox{ if and only if } (\innerl{X}{\psi}-\innerl{m_1}{\psi} )^2-\innerl{X}{\psi}^2<0.\label{centroid-asymp}
\end{equation}

Now, a more precise summary of the above discussion is as follows.

\begin{theorem}\label{th:near}\citep[Th.1]{del12}.
	Let us consider the binary classification problem (\ref{modelogeneral}) under the Gaussian homoscedastic model with $m_0(t)=0$ and continuous $K$. 
	\begin{itemize}
		\item[(a)] If $\sum_{j\geq 1}\theta^{-1}_j \mu_j^2<\infty$, the minimal (Bayes) misclassification probability is given by $err_0=1-\Phi\left(\frac{1}{2}(\sum_{j\geq 1}\theta^{-1}_j \mu_j^2)^{1/2}\right)$. Moreover, under the extra assumption $\sum_{j\geq 1}\theta^{-2}_j \mu_j^2<\infty$, the optimal classifier (that achieves this error) is the rule $T^0$ defined in (\ref{centroid-asymp}).
		\item[(b)] If $\sum_{j\geq 1}\theta^{-1}_j \mu_j^2=\infty$, the minimal misclassification probability is $err_0=0$ and it is achieved, in the limit, by a sequence of classifiers constructed from $T^0$ by replacing the function $\psi$ with 
		$\psi^{(r)}=\sum _{j=1}^r \theta_j^{-1}\mu_j\phi_j(t)$, with $r\uparrow\infty$.
	\end{itemize} 
\end{theorem}


As pointed out in \cite{del12}, 
\textit{``We argue that those [functional classification] problems have unusual, and fascinating, properties that set them apart from their finite dimensional counterparts. In particular we show that, in many quite standard settings, the performance of simple [linear] classifiers constructed from training samples
	becomes perfect as the sizes of those samples diverge [...]. That property never holds for finite dimensional data, except in pathological
	cases.''} 

Our purpose here is to show that the setup of Theorem \ref{th:near} (that is, Theorem 1 in \cite{del12}) can be analysed from the point of view of RKHS theory. We do this in Theorems \ref{th:near-bis} and \ref{th:near-tri} below.

\begin{theorem}
	\label{th:near-bis}
	In the framework of the classification problem considered in Theorem \ref{th:near}, with continuous trajectories and continuous common covariance function $K$, we have
	\begin{itemize}
		\item[(a)] $\sum_{j\geq 1}\theta^{-1}_j \mu_j^2<\infty$ if and only if $P_1\sim P_0$. In that case, the Bayes  rule $g^*$ is 
		\begin{equation}
		g^*(X)=1 \mbox{ if and only if } \innerk{X}{m}-\frac{1}{2}\parallel m \parallel_K^2  >0,\label{equiv-rule}
		\end{equation}
		with the notation of Equation (\ref{eq:BayesParzen}). 
		 The corresponding 
		optimal (Bayes) classification error is 
		$L^*=1-\Phi\left(\normk{m} / 2 \right)$.
		Under the additional condition $\sum_{j\geq 1}\theta^{-2}_j \mu_j^2<\infty$, the optimal rule given in Theorem \ref{th:near} (a) provides an alternative expression of (\ref{equiv-rule}) based on the ``coordinates'' $\theta_j$ and $\mu_j$.
		\item[(b)] $\sum_{j\geq 1}\theta^{-1}_j \mu_j^2=\infty$ if and only if $P_1\bot P_0$.
		In this case the Bayes error is $L^*=0$.
	\end{itemize} 
\end{theorem}

We next make explicit the meaning of the near perfect classification phenomenon. The next theorem establishes that
in the singular case (where the Bayes error is zero) we can  construct a classification rule 
	whose misclassification probability is arbitrarily small. 

\begin{theorem}
	\label{th:near-tri}
Let us consider the singular case analyzed in Theorem \ref{th:near-bis}. Then, there is a sequence of approximating classification problems,  of type $P_{0n}$ vs. $P_{1n}$, corresponding the absolutely continuous case $P_{0n}\sim P_{1n}$, such that $P_{in}$ converges weakly to $P_i$, for $i=0,1$ as $n\to \infty$ and the misclassification probabilities of the respective optimal rules (which are explicitly known) tend to zero.
\end{theorem}

Now, we are in position to comment the contributions of the above Theorems \ref{th:near-bis} and \ref{th:near-tri}, from the perspective of Theorem 1 in \cite{del12} (see Theorem \ref{th:near} above for a slightly simplified version). First, Theorem \ref{th:near-bis} is, in some sense, analogue to the Delaigle-Hall's  result. In the absolutely continuous case, Theorem \ref{th:near-bis} (a) provides a completely general, coordinate-free expression for the Bayes rule. It only requires the condition $\sum_{j\geq 1}\theta^{-1}_j \mu_j^2<\infty$ which is minimal in the sense that it amounts to $P_0\sim P_1$. Moreover, under the Delaigle-Hall's  assumption $\sum_{j\geq 1}\theta^{-2}_j \mu_j^2<\infty$, such Bayes rule can be expressed in ``elementary terms'' with no resort to the stochastic integral $\langle X,m\rangle_K$ which appears in (\ref{equiv-rule}). This highlights an interesting contribution of Theorem 1 in \cite{del12} which remains ``hidden'' unless the whole problem is considered from the RKHS point of view.

Theorem \ref{th:near-bis} (b) and \ref{th:near-tri} shed some light on the ``near-perfect'' classification phenomenon in two specific aspects. First, Theorem \ref{th:near-bis} (b)  shows that Delaigle-Hall`s condition $\sum_{j\geq 1}\theta^{-1}_j \mu_j^2=\infty$ has a probabilistic interpretation in terms of mutual singularity of measures. Second, Theorem \ref{th:near-tri} shows that the classification problem in this singular case can be arbitrarily approximated by a  sequence of problems in the absolutely continuous case for which the Bayes rules are explicitly known. This establishes an useful link between the dual cases of singularity and absolutely continuity.

\section{A model-based proposal for variable selection and classification}
\label{sec:variableselection}

Variable selection methods are quite appealing when classifying functional data since they help reduce noise and remove irrelevant information. Classification performance often improves if we only use their the functional data values at carefully selected points, instead of employing the whole  trajectories.

In this section we argue that the RKHS framework offers a natural setting to formalize variable selection problems.  The ability of RKHS  to deal with these problems is mainly due to the fact that, by the reproducing property, the elementary functions $K(\cdot,t)$ act as a sort of Dirac's deltas. By contrast, the usual $L^2[0,T]$ space lacks  functions   playing a similar role. Thus, we propose a RKHS-based variable selection method which is motivated by  the expressions of Radon-Nikodym derivatives and optimal rules we have derived in the previous sections. We will also see that  our method for identifying the relevant points has an associated classification rule which is consistent under some simple assumptions.

\subsection{The proposed method}\label{subsec:rkvs}

We deal here with the functional supervised classification problem under the model (\ref{modelogeneral}), assuming that the error processes $\epsilon_0$ and $\epsilon_1$ are Gaussian and homoscedastic. If we are willing to use a variable selection methodology, our aim would be to choose suitable, informative enough, points $t_1,\ldots,t_d$  in order to
perform the classification task using just the $d$-dimensional marginal $(X(t_1),\ldots,X(t_d))$. Assume in principle that $d$ is fixed. Then, the natural question is: what is the optimal choice $(t_1^*,\ldots,t_d^*)$ for $(t_1,\ldots,t_d)$?

The answer is simple if we note that under the assumed model the conditional distributions
$(X(t_1),\ldots,X(t_d))|Y=i$, for  $i=0,1$, are Gaussian and homoscedastic with a common covariance matrix whose $i,j$ entry is $K(t_i,t_j)$. Let us denote by $K_{t_1,\ldots,t_d}$ such covariance matrix. Thus, after variable selection, the classification task based on $(X(t_1),\ldots,X(t_d))$ boils down to a standard $d$-variate discrimination problem between two $d$-variate normal populations. Let us denote 
by $m_{t_1,\ldots,t_d}=(m_1(t_1),\ldots,m_1(t_d))-(m_0(t_1),\ldots,m_0(t_d))$ the difference between both mean vectors. It is well-known \citep[p. 244]{ize08} that the optimal misclassification probability (Bayes error) in such a classification problem is a decreasing function of the Mahalanobis distance 
between both mean vectors, $m_{t_1,\ldots,t_d}^\top K^{-1}_{t_1,\ldots,t_d}m_{t_1,\ldots,t_d}$, where $u^\top$ denotes the transpose of $u$. 
 As a consequence, the criterion for variable selection follows in a natural way: we should choose $(t_1^*,\ldots,t_d^*)$ maximizing  $m_{t_1,\ldots,t_d}^\top K^{-1}_{t_1,\ldots,t_d}m_{t_1,\ldots,t_d}$ over a suitable domain.

The theoretical results in this section hold when  we look for the maximum within a compact domain $\Theta\subset [0,T]^d$ such that   $K_{t_1,\ldots,t_d}$ is nonsingular for all $(t_1,\ldots,t_d)\in \Theta$ (so that $K^{-1}_{t_1,\ldots,t_d}$ makes sense). For example, given $\delta>0$, the domain 
	\[
	\Theta=\Theta(\delta)=\{(t_1,\ldots,t_d)\in [0,T]^d:\, t_{(i)}+\delta \leq t_{(i+1)}, \ \mbox{for}\ i=0,\ldots,d-1\},
	\]
	where $t_{(i)}$, $i=1,\ldots,d$, denote the ordered values (with $t_{(0)}:=0$) that fulfil the required conditions if the finite-dimensional distributions of the process $X$ are not degenerated. The value of  $\delta$ can be chosen as small as desired so that the restriction to $\Theta(\delta)$ is not relevant in practice when we can observe the trajectories at a dense enough sample of points.

Denote $\psi(t_1,\ldots,t_d):=m_{t_1,\ldots,t_d}^\top K^{-1}_{t_1,\ldots,t_d}m_{t_1,\ldots,t_d}$. Our criterion for variable selection is the following:  choose $(t_1^*,\ldots,t_d^*)\in \Theta$ such that 
	\begin{equation}
	\psi(t_1^*,\ldots,t_d^*) \geq \psi(t_1,\ldots,t_d),\ \mbox{for all}\ (t_1,\ldots, t_d) \in \Theta. \label{eq:rkvs}
	\end{equation}
Since $m$ and $K$ are usually unknown, we propose to replace them by appropriate estimators  $\hat{m}_{t_1,\ldots,t_d}(t)$  and $\hat K_{t_1,\ldots,t_d}$ (more on this below).  
	The criterion we suggest for  variable selection in practice   is to choose   points
	$(\hat{t}_1,\ldots,\hat t_d\ )\in \Theta$ such that $
	\hat \psi(\hat t_1,\ldots, \hat t_d) \geq \hat \psi(t_1,\ldots, t_d)$  for all $(t_1,\ldots, t_d)\in \Theta$, where 
	\begin{align}
	\hat \psi(t_1,\ldots,t_d):= \hat{m}_{t_1,\ldots,t_d}^\top \hat K_{t_1,\ldots,t_d}^{-1} \hat{m}_{t_1,\ldots,t_d}.\label{eq:finalvs}
	\end{align}
We will denote this variable selection method by RK-VS (RK comes from ``reproducing kernel'', in view of the RKHS interpretation we will give in subsection \ref{subsec:explanation}). 

\

\textit{ On the estimation of $m$ and $K$}. In principle (unless some strong parametric assumptions are made), the estimation of $m=m_1-m_0$ will be done in the simplest way, using the sample means, i.e., 
	$\hat m=\hat{m}_1(t) - \hat{m}_0(t)$, where
	$
	\hat{m}_j(t) := n_j^{-1} \sum_{i=1}^{n_j}X_{j,i}(t)  = \bar{X}_j(t),
	$
	for $j=0,1$.

The estimation of $\hat K$ might look as a more delicate issue. It is well-known that in some functional data analysis techniques (including functional linear regression and principal components analysis) there is a need to use smooth estimators of the covariance operator $K$; see, for example, \citet[Secs. 5.2 and 7.1]{cue14}. Of course, such smoothed estimators could also be applied here  but the underlying (functional) reasons to use them are not present in this case since in fact we are only concerned with the covariance matrices $K_{t_1,\ldots,t_d}$ of finite dimensional projections
	$(X(t_1),\ldots,X(t_d))$. Thus, unless otherwise stated, we will estimate $K_{t_1,\ldots,t_d}$ by the natural empirical counterpart  $\hat K_{t_1,\ldots,t_d}$ constructed from the sample covariances. This has been the method we have used (with overall good results) in our empirical studies. Again a natural alternative to such estimators would arise in those cases in which we are assuming a precise parametric model, such as for example a Brownian motion
	for which $K(s,t)=K(\theta,s,t)=\theta\, \mbox{min}(s,t)$ depending on an unknown parameter $\theta$. In such models one could naturally consider parametric estimations of type $K(\hat \theta,s,t)$.

Some further practical issues associated with the use of the RK-VS method will be considered below in Subsection \ref{sub:prac}. Before that, we are going to study the functional interpretation of these methodology.

\subsection{An interpretation in functional terms}\label{subsec:explanation}
Let us focus again on the homoscedastic Gaussian functional classification problem (\ref{modelogeneral})  in the absolutely continuous case. According to Theorem \ref{theorem:parzen}, $P_0\sim P_1$ entails $m\in{\mathcal H}(K)$, where $m=m_1-m_0$. Now assume that 

\begin{equation}
m(\cdot)=\sum_{i=1}^d \alpha_i K(\cdot,t_i),\label{mm}
\end{equation}
for some $d\in{\mathbb N}$, $\alpha_i\in {\mathbb R}$, $t_i\in[0,T]$, $i=1,\ldots,d$. Note that, for $d$ large enough, a finite linear combination of type (\ref{mm}) would be a good approximation for the true value of $m$. This makes sense since, from the definition of the RKHS space ${\mathcal H}(K)$, the set ${\mathcal H}_0(K)$ of such finite linear combinations is dense in ${\mathcal H}(K)$. So the homoscedastic classification problem (\ref{modelogeneral}) with $m\in{\mathcal H}_0(K)$ can be seen as an approximation to the general problem with $m\in{\mathcal H}(K)$. 

Let us now recall that, from Theorem   \ref{theorem:optParzen}, the optimal rule to classify a trajectory $x$ between $P_0$ and $P_1$ (with $P_0\sim P_1$) is  $g^*(x)=\mathbb{I}_{\{\eta^*(x)>0\}}$, where $\eta^*(x)$ is given in Equation (\ref{eq:BayesParzen}). If $m$ has the form indicated in (\ref{mm}), the discriminant score $\eta^*(x)$ is given by
\begin{align*}
\eta^*(x) =&  \innerk{x-m_0}{\sum_{i=1}^d \alpha_i K(\cdot,t_i)} - \frac{1}{2}\normk{\sum_{i=1}^d \alpha_i K(\cdot,t_i)}^2 - \log\left(\frac{1-p}{p}\right)\\
=&  \sum_{i=1}^d \alpha_i (x(t_j)-m_0(t_j)) - \frac{1}{2}\sum_{i=1}^d\sum_{j=1}^d\alpha_i\alpha_j K(t_i,t_j)- \log\left(\frac{1-p}{p}\right),
\end{align*}
where we have used the reproducing property to obtain the last equality. 

A more familiar expression for the optimal rule is obtained taking into account that 
(\ref{mm}) implies the following relationship between  $\alpha_1,\ldots,\alpha_d$ and  $t_1,\ldots,t_d$:
\begin{equation}
\label{eq:alpha}
m_{t_1,\ldots,t_d} = K_{t_1,\ldots,t_d} \cdot (\alpha_1,\ldots,  \alpha_d)^\top,
\end{equation}
Now, using (\ref{eq:alpha}) we can write
\begin{equation}
\label{eq:BayesRule}
\eta^*(x) = \sum_{i=1}^d \alpha_i\left(x(t_i)-\frac{m_0(t_i)+m_1(t_i)}{2}\right)- \log\left(\frac{1-p}{p}\right),
\end{equation}
which exactly coincides with the discriminant score of the optimal (Bayes) rule for the finite dimensional discrimination problem based on the $d$-dimensional marginals 
$(X(t_1),\ldots,X(t_d))$. 

Note also that if $m$ is given by (\ref{mm}) then 
\[
\|m\|^2_K = \sum_{i=1}^d\sum_{j=1}^d\alpha_i\alpha_jK(t_i,t_j) =  m_{t_1,\ldots,t_d}^\top K_{t_1,\ldots,t_d}^{-1} m_{t_1,\ldots,t_d} 
\]

We now summarize the previous discussion in the following statement.

\begin{proposition}\label{prop:explanation}
	Let us consider the  functional classification problem 
	of discriminating between the processes $P_0$ and $P_1$ with continuous mean functions $m_i$ and continuous trajectories of type $X(t):=m_i(t)+\epsilon_i(t)$, $t\in[0,T]$, where the $\epsilon_i$ are independent Gaussian non-degenerate processes with mean 0 and common continuous covariance function $K(s,t)$. Then,
	
	(a) the $d$-dimensional classification problem of discriminating between $P_0$ and $P_1$ on the sole basis of the projections $(X(t_1),\ldots,X(t_d))$ at given points $t_1,\ldots,t_d$ is equivalent (in the sense of having the same optimal rule and Bayes error) to the functional problem stated in the previous paragraph whenever $m:=m_1-m_0$ has the form 
	$m(\cdot)=\sum_{i=1}^d\alpha_iK(\cdot,t_i)$.

	(b) Denote by $K_{t_1,\ldots,t_d}$ the covariance matrix
	of $(X(t_1),\ldots,X(t_d))$ and let $m_{t_1,\ldots,t_d}$ be the difference between both mean vectors. The Mahalanobis distance between the distributions $(X(t_1),\ldots,$ $X(t_d))|Y=i$ for $i=0,1$, given by $m^\top_{t_1,\ldots,t_d} K_{t_1,\ldots,t_d}^{-1} m_{t_1,\ldots,t_d}$ coincides with $\Vert m\Vert_K^2$, the norm of $m$ in the RKHS induced by $K$, provided again that $m(\cdot)=\sum_{i=1}^d\alpha_iK(\cdot,t_i)$.
		
	(c) The optimal choice for $(t_1,\ldots,t_d)$, in the sense of minimizing the classification error, is obtained by maximizing
	$\Vert m\Vert_K^2$ among all functions $m$ in the RKHS space having an expression of type $m(\cdot)=\sum_{i=1}^d\alpha_iK(\cdot,t_i)$.
\end{proposition}

At this point, one might wonder about the role of the assumption $m(\cdot)=\sum_{i=1}^d\alpha_iK(\cdot,t_i)$. The natural question is: to what extent such condition is needed in our approach to variable selection? In this respect, it is particularly important to note that the method defined in (\ref{eq:finalvs}), \textit{still makes sense even if such assumption is not fulfilled}; in that case, the method provides (asymptotically) the best choice $(X(t_1^*),\ldots,X(t_d^*))$ of the chosen number $d$ of variables in order to obtain a maximal separation in the Mahalanobis distance for their mean vectors under $P_0$ and $P_1$. Note that, in principle, this idea could be considered without any assumption on the functional model (except, perhaps, homoscedasticity). The contribution of Proposition \ref{prop:explanation} is just to establish in precise terms the conditions on the \textit{functional} classification model under which the proposed variable selection procedure will be (asymptotically) optimal; see Theorem \ref{theorem:fisher} below.

\subsection{The RK-based classification rule: consistency}\label{sub:consistency}
The above described RK-VS variable selection method has an associated classification rule  which is just the classical Fisher's linear rule for the discrimination problem based on the RK-VS selected variables $(X(\hat t_1),\ldots,X(\hat t_d))$. This classifier will be denoted RK-C.

The following result shows that the estimation procedure in the definition of the RK-C method works, in the sense that the performance of the classification procedure implemented with the variables corresponding to the estimated points $\hat t_1,\ldots,\hat t_d$ tends, as the sample size increases, to that achieved with the optimal points 
$t_1^*,\ldots,t_d^*$ defined in equation (\ref{eq:rkvs}). 
This is next formalized. Let us consider   again our functional supervised classification problem under the conditions stated in the first paragraph of Proposition \ref{prop:explanation}. Let $L^*=\mathbb{P}(g^*(X)\neq Y)$ be the  misclassification probability obtained with the RK-C classifier, when both $m$ and $K$ are known and we use the ``ideal'' variable selection choice $(X(t_1^*),\ldots,X(t_d^*))$.
	Denote by $L_n=\mathbb{P}(\hat{g}(X)\neq Y|X_1,\ldots,X_n)$  the misclassification probabilities  of Fisher's rules defined in terms
	of $(X(\hat t_1),\ldots,X(\hat t_d))$ (see Equations (\ref{eq:rkvs}) and (\ref{eq:finalvs}) above). For the sake of simplicity consider $p=1/2$. In this setup we have the following consistency result under fairly general conditions:

\begin{theorem}
	\label{theorem:fisher}
	Consider the classification problem (with $p=1/2$) according to the model (\ref{modelogeneral}), for $t\in[0,T]$. Denote
	$\hat{m}(t) = \hat{m}_1(t) - \hat{m}_0(t)$, where $\hat{m}_j(t) := n_j^{-1} \sum_{i=1}^{n_j}X_{j,i}(t)  = \bar{X}_j(t)$ for $j=0,1$, and let $\hat{K}_{t_1,\ldots,t_d}$ be the pooled sample covariance matrix, whose $(i,j)$ entry is 
	$
	\hat{K}_{t_1,\ldots,t_d}(i,j) = \sum_{r\in\{0,1\}} \left(\frac{1}{n_r}\sum_{\ell =1}^{n_r}(X_{r,\ell}(t_i)-\bar{X}_r(t_i))(X_{r,\ell}(t_j)-\bar{X}_r(t_j))\right).
	$
	
	Assume,	
	\begin{itemize}
		\item[(i)] $\mathbb{E}\|\epsilon_j^2\|_\infty<\infty$, for $j=0,1$, where $\|\cdot\|_\infty$ stands for the supremum norm.
		\item[(ii)] The variable selection method is performed on a compact set $\Theta\subset[0,T]^d$. 
		\item[(iii)] ${K}_{t_1,\ldots,t_d}$ is invertible for all  $(t_1,\ldots,t_d)\in\Theta$ and their entries
		are continuous on $\Theta$. 
	\end{itemize}	
	Then, $L_n\to L^*$ a.s., as $n\to\infty$.

\end{theorem}

Note that when the mean difference has the form $m(\cdot)=\sum_{i=1}^d\alpha_iK(\cdot,t_i)$, and $(t_1,\ldots,t_d)\in \Theta$,  from Proposition \ref{prop:explanation} we have that $L^*$ in Theorem \ref{theorem:fisher} coincides in fact with the Bayes error in the original functional problem. Also, assumption (ii) entails that $t_1<\ldots<t_d$ for all $(t_1,\ldots,t_d)\in \Theta$.
Note finally that the same result would still be valid for other estimators of $m$ and $K$ as long as they are consistent uniformly on $\Theta$  (see the proof of Theorem \ref{theorem:fisher} in the Supplementary Material document). This will be typically the case when we may assume that the covariance operator is indexed by (and depends continuously on) a finite-dimensional parameter $\theta$, so that we only need to estimate $\theta$.

\subsection{Some practical issues and computational aspects}\label{sub:prac}

There are several aspects worth of attention in the RK-VS and RK-C procedures, as presented in the previous subsections. 

First, the number $d$ of points to be selected is assumed to be finite. This can be seen as a reasonable approximation since, as mentioned above, the set of all finite linear combinations $\sum_i\alpha_iK(\cdot,t_i)$ is dense in the RKHS space ${\mathcal H}(K)$ to which $m$ is assumed to belong. Also, in many practical situations, the mean function $m$ depends  just on a finite number of values $t_i$. A simple example of this situation is as follows: consider model (\ref{modelogeneral}) where $\epsilon_0$ and $\epsilon_1$ are Brownian motions, $m_0\equiv 0$ and $m_1$ is a continuous, piecewise linear function such that $m_1(0)=0$. According to the computations above, the discriminant score of a trajectory $x(t)$ only depends  on the values of $x$ at the points where $m_1$ is not differentiable (and, possibly, also on $x(0)$ and $x(T)$). This can be more easily derived from the representation of the discriminant scores in terms of stochastic integrals (see Subsection \ref{sub:parzen}, remark~(b)).  

Second, the matrix $K_{t_1,\ldots,t_d}$ and the prior probability $p$ may not be known either. Thus, $K_{t_1,\ldots,t_d}$ and $p$ might be replaced by  suitable consistent estimators $\hat K_{t_1,\ldots,t_d}$ and $\hat p$. The appropriate estimator  $\hat K_{t_1,\ldots,t_d}$ depends on the assumptions we are willing to make about the processes involved in the classification problem. For instance, if all we want to assume is that they are Gaussian, we could use the pooled sample covariance matrix. However, under a parametric model, only a few parameters should be estimated in order to get $\hat K_{t_1,\ldots,t_d}$; see Subsection \ref{sub:motiv} for more details on this.

Third, $\hat \psi(t_1,\ldots,t_d)$ is a non-concave function with potentially many local maxima so that the maximization process could be hard to implement even for moderately large values of $d$. 
Hence, in practice, we can use the following ``greedy'' algorithm.

\begin{enumerate}
	\item Initial step: consider a large enough grid of points in $[0,T]$ and find $\hat t_1$  such that $\hat \psi(\hat{t}_1)\geq \hat \psi(t)$ when $t$ ranges over the grid. Observe that this initial step amounts to find the point maximizing the signal-to-noise ratio since
	\[
	\hat \psi(t) = \frac{\hat{m}(t)^2}{\hat \sigma^2_t} = \frac{(\bar{X}_1(t) -\bar{X}_0(t))^2}{\hat \sigma^2_t},
	\]
	for a suitable estimator $\hat \sigma^2_t$ of the variance at $t$.
	
	\item Repeat until convergence: once we have computed $\hat t_1,\ldots, \hat t_{d-1}$, find $\hat t_{d}$  such that \\ $\hat \psi(\hat t_1,\ldots, \hat t_{d-1}, \hat t_d)\geq~\hat\psi(\hat t_1,\ldots, \hat t_{d-1}, t)$ for all $t$ in the grid.
\end{enumerate}

Whereas we have no guarantee that this algorithm converges to the global maximum of $\hat \psi(t_1,\ldots,t_d)$, it is computationally affordable and shows  good performance in practice.

\subsection{An illustrative example. The price of estimating the covariance dunction} \label{sub:motiv}

The purpose of this subsection is to gain some practical insight 
on the meaning and performance of our RK methods. In particular, 
we will take into account that
the RK methods can incorporate information on the assumed underlying model, via a known (or partially known) covariance function. In what follows we will assume that the data trajectories come from a Brownian Motion with different (unknown) mean functions. So we would incorporate this information in our ``variable selection + classification'' task by just  using the, supposedly true, $K(s,t)$, instead of its estimator in (\ref{eq:finalvs}). We will denote by RK$_{B}$-VS and RK$_{B}$-C the resulting ``oracle'' methods for variable selection and classification, respectively, implemented with $K(s,t)=\min\{s,t\}$.

Of course, the assumption that $K$ is known is too strong, but still it is useful to compare the performance of the  oracle RK$_{B}$-VS and RK$_{B}$-C methods with the standard RK-VS and RK-C versions in which $K(s,t)$ is estimated from the sample. In particular, we want to assess  the loss of efficiency involved in the estimation of $K(s,t)$.  To this end, consider a simulated example under the general model  (\ref{modelogeneral}) in which $P_0$ and $P_1$ are Brownian motions whose mean functions fulfil $m(t)=m_1(t)-m_0(t)=\sum_{i=1}^r a_i\Phi_{m,k}(t)$, where $t\in[0,1]$, the $a_i$ are constants and the $\{\Phi_{m,k}\}$ are continuous piecewise linear functions as those considered in \citet[p. 28]{mor10};
they are obtained by integrating the piecewise constant functions of a Haar basis. Explicit expressions can be found in the Supplementary Material document.
 In fact, it can be proved there that the $\{\Phi_{m,k}\}$ form a orthonormal basis of the Dirichlet space $\mathcal{D}[0,1]$ which, as commented above, is the RKHS space corresponding to this model. As a consequence, the equivalence condition in Theorem \ref{theorem:optParzen} is automatically fulfilled. In addition, given the simple structure of the ``peak'' functions $\Phi_{m,k}$, it is easy to see that the ``sparsity condition'' $m(\cdot)=\sum_{i=1}^d\alpha_i K(\cdot,t_i)$ also holds in this case.  
To be more specific, in our simulation experiments we have taken  $m_0(t)=0$, $m_1(t)=\Phi_{1,1}(t)-\Phi_{2,1}(t) +\Phi_{2,2}(t) - \Phi_{3,2}(t)$,  and $p=\mathbb{P}(Y=1)=1/2$, so that the Bayes rule given by Theorem
\ref{theorem:optParzen} depends only on the values $x(t)$ at $t$ = 0, 1/4, 3/8, 1/2, 3/4 y 1 and the Bayes error is 0.1587. Some typical trajectories are shown in Figure S1 in the Supplementary Material document. 

\begin{figure}
	\centering
	\begin{minipage}{0.45\textwidth}
		\centering 		 
		\includegraphics[width=1.1\linewidth]{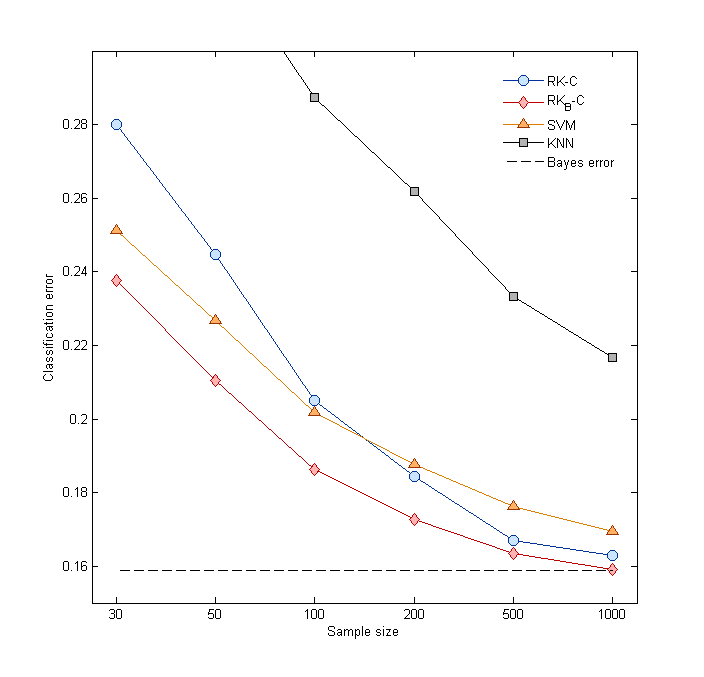}
	\end{minipage}\hfill
	\begin{minipage}{0.45\textwidth}
		\centering
		\includegraphics[width=0.95\linewidth]{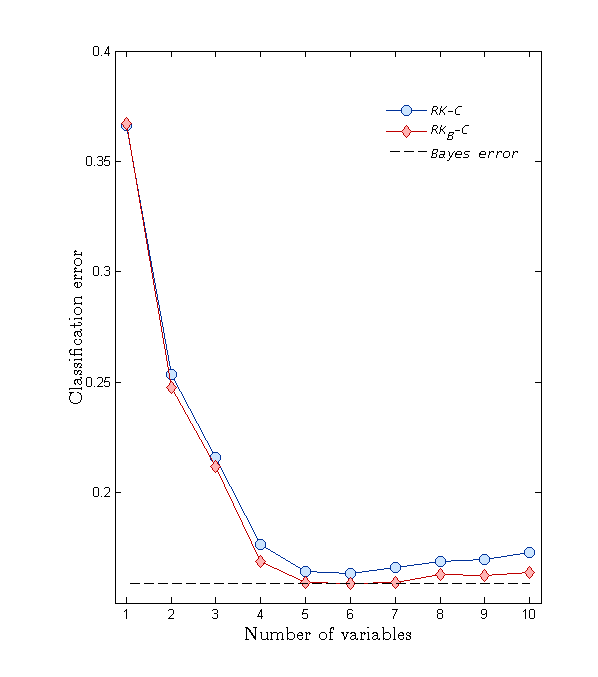} 		
	\end{minipage}
	\caption{\footnotesize Evolution of the classification error of RK-C and RK$_{B}$-C in terms of the sample size (left panel) and the number of selected variables (right panel).}\label{fig:acc}
\end{figure}

\begin{figure}
	\centering
	\includegraphics[width=1\linewidth]{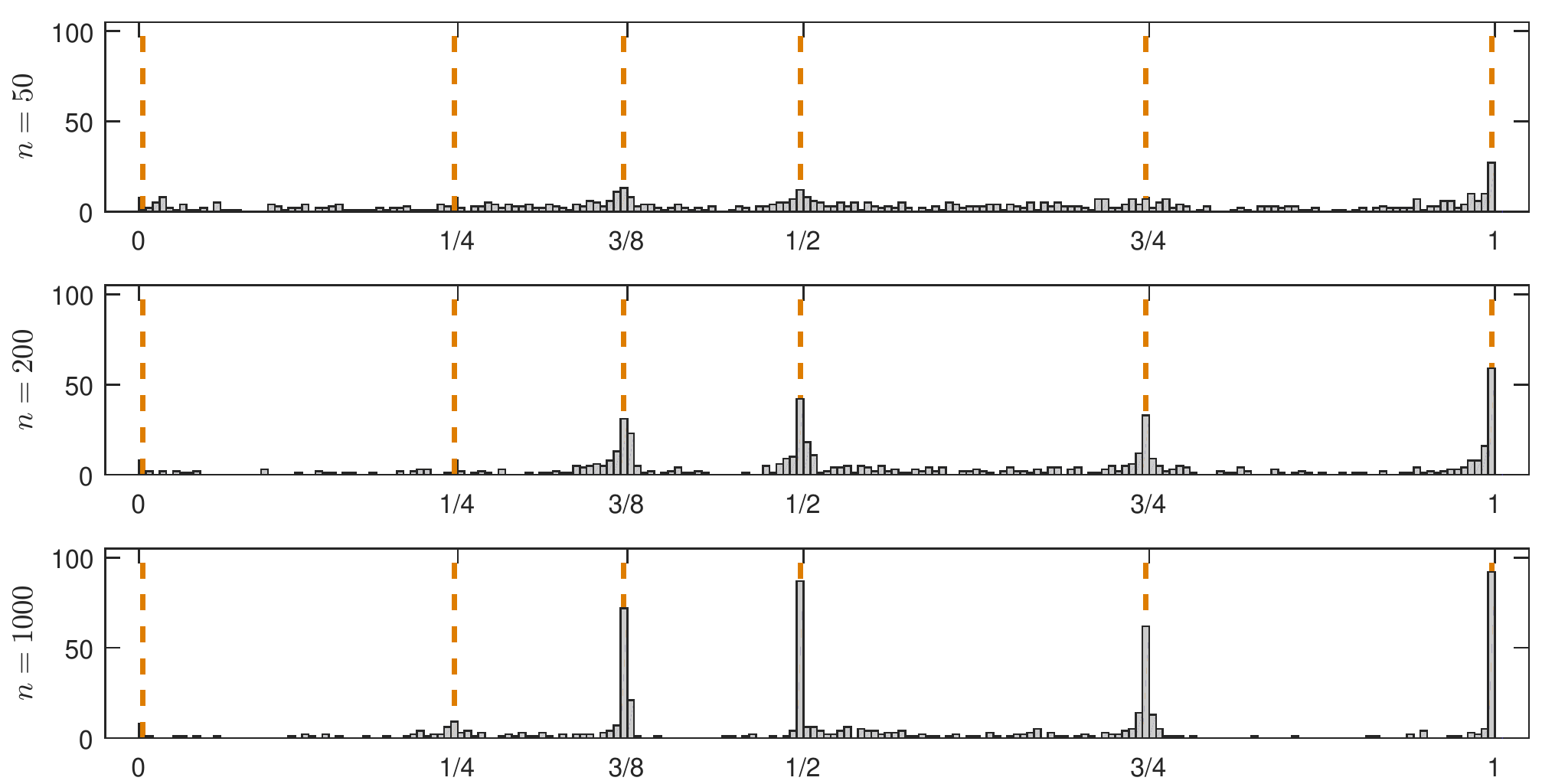} 	
	\caption{\footnotesize Histograms of the six first selected variables by RK-VS over 100 runs for sample sizes 50 (top panel), 200 (middle panel) and 1000 (bottom panel).}\label{fig:hist}
\end{figure}

Now, we analyze the performance of RK and RK$_{B}$ in this example.  The left panel of Figure \ref{fig:acc}  shows the evolution of the classification error as the sample size increases for  RK-C (blue line with circles), RK$_{B}$-C (red line with diamonds), $k$-nearest neighbor rule (kNN, gray line with squares)  and the support vector machine classifier with a linear kernel (SVM, orange line with triangles). The last two rules are applied to the complete trajectories, without any variable selection. The dashed black line indicates the Bayes error. Each output is obtained by averaging 100 independent runs with test samples of size 200; for each sample size, the number of selected variables (RK-C and RK$_{B}$-C), the number $k$ of neighbours (kNN) and the cost parameter (SVM) are set through a validation sample. The right panel of Figure \ref{fig:acc} shows the classification error in terms of the number of variables for RK-C and RK$_{B}$-C for $n=500$. Finally, Figure \ref{fig:hist} shows the frequency of selection of each variable among the first six (by construction, we know there are just six relevant points) corresponding to 100 independent runs of RK-VS for  three different sample sizes. The theoretical relevant points are marked by vertical dashed lines.  So, to sum up, whereas Figure \ref{fig:acc} summarizes the results in terms of classification performance, Figure \ref{fig:hist} is more concerned with capacity of identifying the right relevant variables.

These results are quite positive; RK-C seems to be a good estimator of the optimal classifier as the error rate converges swiftly to the Bayes error even when the number of variables is unknown and fixed by validation. Observe that the convergence seems to be slower for other standard classifiers such as kNN and SVM (Figure \ref{fig:acc}, left plot). The right plot in Figure \ref{fig:acc} shows that for the true number of variables (six) the algorithm achieves the best performance. By contrast, a wrong choice of the number of variables can entail an important increase of the misclassification rate, so this is a sensitive issue. In addition, the selected variables (represented in Figure \ref{fig:hist}) are mostly in coincidence with the theoretical ones. Even for small sample sizes, RK$_{B}$-VS and RK-VS variables are grouped around the relevant variables. Only the variable $X(0)$  is omitted since it is in fact nearly irrelevant. This good performance in detecting 
the important variables is in principle better than one might expect for a greedy algorithm (that, therefore might not provide the true global optimum).  Note also that the inclusion of some additional information seems specially beneficial for smaller sample sizes. Finally, it is worth mentioning that the RK-based methods seem to be relatively inexpensive from the computational point of view. For example, the increase in the computation time as the sample size increases is much slower than that of other competing methods. See Figure S2 in the Supplementary Material document.

\section{Experiments}\label{sec:experiments}

Our purpose in Section \ref{sec:variableselection} was twofold: we proposed both a variable selection method and an associated classifier. We check here the corresponding performances.

\subsection{Simulation study}\label{sub:simulation}

The simulation experiments include  94 models,  previously considered in the studies by \cite{ber15a,ber15b}. These models can be grouped into three classes.

(i) \textit{Gaussian models}: they are defined via the marginal Gaussian distributions (Brownian-like, Ornstein Uhlenbeck,...)  $P_i$ of $X(t)|Y=i$ for $i=0,1$. In all cases $p={\mathbb P}(Y=1)=1/2$.

(ii) \textit{Logistic-type models}: they are defined through the function $\eta(X)={\mathbb P}(Y=1|X(t))$ and the marginal of $X$. It is assumed that  $\eta(x)=(1+e^{-\psi(x(t_1),\cdots,x(t_d))})^{-1}$,
with different choices for the link function $\Psi$.

(iii) \textit{Finite mixtures} of different types of Gaussian models. 

Detailed descriptions of the 94 considered models can be found in the Supplementary Material document. We should emphasize that only 7 among these 94 models fulfill all the conditions imposed in our theoretical results. They are grouped under the label RKHS in the output tables of the Supplementary Material document. The remaining ``unorthodox'' models aim at checking the behavior of our proposal when some departures from the assumptions are present.

Training samples of sizes $n=30,50,100,200$ are considered for each model. Sample trajectories are discretized in 100 equispaced points in the interval [0,1]. The criterion of comparison is the classification accuracy for an independent test sample of size 200. The number of selected variables as well as the classification parameters (if needed) are fixed in a validation step, using, for each test sample, another 
independent validation sample of size 200.  The final output is the average classification accuracy over 200 runs of this experiment. 

\vspace{0.4cm}

\noindent \it Comparison of variable selection methods\rm

The primary aim of the study is to check the performance of our RK variable selection method against other dimension reduction procedures, chosen among the winners in  \cite{ber15a,ber15b}. To be specific, these are the methods considered in the experiments: 
\begin{itemize}
	\item RK-VS, as defined in (\ref{eq:finalvs}).
	\item RK$_{B}$-VS, the ``oracle'' version RK-VS defined in Subsection \ref{sub:motiv} by assuming that the common covariance structure coincides with that of the Brownian motion. Since this is not in general a realistic assumption, RK$_{B}$ is included only for illustration purposes, just to check the price of the estimation in $K(s,t)$ and the (sometimes surprising) resistance  against the assumptions on the covariance structure.
	
	\item  mRMR-RD: this is a modified version of the popular minimum redundancy maximum relevance algorithm (mRMR) for variable selection  proposed by \cite{din05}. The aim of mRMR is to select the subset $S$ of variables that maximizes the difference $\mbox{rel}(S)-\mbox{red}(S)$, where $\mbox{rel}(\cdot)$ and $\mbox{red}(\cdot)$ are appropriate measures of relevance and redundancy which are defined in terms of an association measure between random variables. The improved version of mRMR considered here (denoted mRMR-RD) has been recently proposed in \cite{ber15a}. It relies on the use of the increasingly popular  \textit{distance correlation} \citep{sze07}
	association measure to define relevance and redundancy in the mRMR algorithm.  
	\item  MHR: the maxima hunting method \citep{ber15b} also uses the distance correlation $R^2(t)=\mathcal{R}^2(X(t),Y)$, between $X(t)$ and the binary response $Y$ to select the points $t_1,\ldots,t_k$ corresponding to the local maxima of $R^2(t)$. This automatically takes into account the relevance-redundancy trade-off (though in a qualitative way, quite different to that of the mRMR methodology).
	\item  PLS: partial least squares, a well-known dimension reduction technique; see e.g. \cite{del12} and references therein. 
\end{itemize}

All these methods for variable selection (or, in the case of PLS, for projection-based dimension reduction) are data-driven, i.e., independent on the classifier, so we can combine them with different classifiers. For illustrative purposes we show the results we have obtained with the Fisher linear classifier (LDA), $k$ nearest neighbors (kNN) and support vector machine with a linear kernel (SVM).

Some aggregated results  are in Table \ref{table:vs}. Variable selection methods and PLS are in columns and each row corresponds to a sample size and a classifier.  Each output is the average classification accuracy of the 94 models over 200 runs.  Boxed outputs denote the best result for each sample size and classifier. The full results of the 1128 experiments (94 models $\times$ 4 samples sizes $\times$ 3 classifiers)  are available in the supplementary file \textit{outputs}. Additional, more detailed, summary tables are included in the Supplementary Material document.

\begin{table}
	\begin{center}
		\caption{\footnotesize Percentage of correct classification  with the three considered classifiers}\label{table:vs}
		
		\medskip
		
		\footnotesize
		\begin{tabular}{llccccc}
			\toprule
			\small Classifier & \small Sample size & \multicolumn{5}{c}{\small Dimension reduction methods}\\ \cmidrule{3-7} 
			& &  mRMR-RD &  PLS &  MHR &  RK-VS &  RK$_{B}$-VS\\
			\midrule
			\small LDA & $n=30$ & 81.04 & \framebox{82.87} & 82.44 & 81.50 & 80.89\\
			& $n=50$ & 82.37 & \framebox{83.78} & 83.68 & 83.44 & 82.54\\
			& $n=100$ & 83.79 & 84.70 & 84.97 & \framebox{85.30} & 84.46\\
			& $n=200$ & 84.88 & 85.46 & 85.90& \framebox{86.51} & 85.90\\
			\midrule
			
			\small kNN & $n=30$ & 81.88 & 82.45 & \framebox{82.46} & 82.28 & 81.92\\
			& $n=50$ & 82.95 & 83.49 & 83.43 & \framebox{83.75} & 83.25\\
			& $n=100$ & 84.31 & 84.77 & 84.73 & \framebox{85.59} & 84.95\\
			& $n=200$ & 85.38 & 85.79 & 85.91 & \framebox{87.16} & 86.50\\
			\midrule
			
			\small SVM & $n=30$ & 83.22 & 84.12 & \framebox{84.62} & 84.28 & 84.12\\
			& $n=50$ & 84.21 & 85.04 & 85.44 & \framebox{85.60} & 85.20\\
			& $n=100$ & 85.27 & 86.03 & 86.29 & \framebox{86.96} & 86.48\\
			& $n=200$ & 86.10 & 86.79 & 86.86 & \framebox{87.90} & 87.50\\
			\bottomrule
			
		\end{tabular}
	\end{center}
\end{table}

The results are quite similar for all considered classifiers: RK-VS  methodology outperforms the other competitors on average with a better performance for bigger sample sizes. Although RK-VS could have more difficulties to estimate the covariance matrix for small sample sizes, it is very close to MHR, which seems to be the winner in that case. Besides, the number of variables selected by RK-VS (not reported here for the sake of brevity; see Table S4 in the Supplementary Material) is comparable to that of mRMR-RD and MHR for kNN and SVM but it is about half of the number selected by mRMR-RD and MHR  for LDA (the number of PLS components is often smaller but they lack  interpretability). Note that, according with the available experimental evidence \citep{ber15b,ber15a}, the competing selected methods (mRMR-RD, MHR and PLS) have themselves a good general performance. So, the outputs in Table \ref{table:vs} are remarkable and encouraging especially taking into account that only 7 out of 94 models under study fulfil all the regularity conditions required for RK-VS. Note  that, somewhat surprisingly, the failure of the ``Brownian assumption'' implicit in the RK$_{B}$-VS method does not entail a big loss of accuracy with respect to the ``non-parametric'' RK-VS version. 

\vspace{0.4cm}

\noindent \it Comparison of classifiers\rm

We also assess the performance of the classifiers RK-C and RK$_{B}$-C;  see the definitions in the first paragraphs of Subsections \ref{sub:consistency} and \ref{sub:motiv}, respectively.  The competitors are kNN and SVM (with linear kernel), two standard all-purpose classification methods. 

Table \ref{table:classify} provides again average percentages of correct classification over 200 runs of the previously considered 94 functional models. The results are grouped by sample size (in rows). Classification methods are in columns. The full detailed outputs are given in the supplementary file \textit{outputs}.

The difference with Table \ref{table:vs}  is that, in this case, the classifiers kNN and SVM are used  with no previous variable selection. So, the original whole functional data are used. This is why we have replaced the standard linear classifier LDA (which cannot be used in high-dimensional or functional settings) with the   LDA-Oracle  method which is just the Fisher linear classifier based on the ``true'' relevant variables (which are known beforehand since we consider models for which the Bayes rule depends only on a finite set of variables). Of course this classifier is not feasible in practice; it is included here only for comparison purposes.

\begin{table}
	\begin{center}\footnotesize 
		\caption{\footnotesize Average classification accuracy (\%) over all considered models.}\label{table:classify}\medskip
		\begin{tabular}{lcccccc}
			\toprule
			\small $n$ & \small kNN & \small SVM & \small RK-C & \small RK$_{B}$-C &  \small LDA-Oracle\\
			\midrule
			$30$ & 79.61& 83.86& 81.50 & 80.89& 84.97\\
			$50$ & 80.96& 85.01& 83.44 & 82.54& 86.23\\
			$100$ & 82.60& 86.20& 85.30 & 84.46& 87.18\\
			$200$ & 83.99& 87.07& 86.51 & 85.90& 87.69\\
			\bottomrule 
		\end{tabular}
	\end{center}
\end{table}

As before, RK-C results are better for higher sample sizes and the distances between SVM or LDA-Oracle and RK-C are swiftly shortened with $n$; and again, RK$_{B}$-C is less accurate than RK-C but not too much. While the global winner is SVM, the slight loss of accuracy associated with the use of RK-C and RK$_{B}$-C
can be seen as a reasonable price for the simplicity and ease of interpretability of these methods. Note also that the associated procedure of variable selection can be seen as a plus of RK-C. In fact, the combination of RK-VS with  SVM outperforms SVM based on the whole functional data.   

\begin{table}
	\begin{center}\footnotesize
		\caption{\footnotesize Average classification accuracy (\%) for the models satisfying the assumptions of Th.~\ref{theorem:fisher}} \label{table:classify2}\medskip
		\begin{tabular}{lccccc}
			\toprule
			\small $n$ & \small kNN & \small SVM & \small RK-C & \small RK$_{B}$-C & \small LDA-Oracle\\
			\midrule
			$30$ & 83.20 & 87.29 & 88.30 & 89.95 & 90.91\\
			$50$ & 84.90 & 88.81 & 89.81 & 90.69 & 91.41\\
			$100$ & 86.61 & 89.88 & 90.81 & 91.18 & 91.64\\
			$200$ & 87.94 & 90.48 & 91.13 & 91.30 & 91.71\\
			\bottomrule
		\end{tabular}
	\end{center}
\end{table}

Table \ref{table:classify2} shows average percentages of correct classification over 200 runs of the subset of models among all seven models  that satisfy the assumptions in Theorem \ref{theorem:fisher}, which establishes the consistency of the procedure proposed in Section \ref{sec:variableselection}. It is not surprising that for these models RK-C and RK$_{B}$-C have a better performance than kNN and SVM. In fact the RK percentages of correct classification are very close to those of LDA-Oracle, which means that there is not much room for improvement under these asumptions.

\subsection{Real data}

We now study the RK-C performance in two real data examples. We have chosen the ``easiest'' and the ``hardest'' data sets (from the classification point of view) of those considered in \cite{del12}.  Given the close connections between our theoretical setting and that of these authors, this partial coincidence of data sets seems pertinent. 

Thus, we follow the same methodology as in the cited paper, that is, we divide the data set randomly in a training sample of size $n$ ($n=30,50,100$) and a test sample with the remaining  observations. Then, the RK-C classifier is constructed from the training set and it is used to classify the test data. The misclassification error rate is estimated through 200 runs of the whole process. The number of variables selected by RK-C is fixed by a standard leave-one-out cross-validation procedure over the training data. 

We consider the \textit{Wheat} and the \textit{Phoneme} data sets. \it Wheat\/ \rm data correspond to 100 near infrared spectra of wheat samples measured from 1100nm to 2500nm in 2nm intervals. Following \cite{del12} we divide the data in two populations according to the protein content (more or less than 15) and use the derivative curves obtained with splines. For this wheat data the near perfect classification is achieved. \it Phoneme \/ \rm is a popular data set in functional data analysis. It consists of log-periodograms obtained from the pronunciation of five different phonemes recorded in 256 equispaced points. We consider the usual binary version of the problem, aimed at classifying the phonemes ``aa'' (695 curves) and ``ao'' (1022 curves). This is not an easy problem.  As in the reference paper we make the trajectories continuous with a local linear smoother and remove the noisiest part keeping the first 50 variables. More details and references on this data can be found 
in \cite{del12}.

Table \ref{table:real2} shows exactly the same results of Table 2 in \cite{del12} plus an extra column (in boldface) for our RK-C method. Since we have followed the same methodology, the results are completely comparable despite the minimum differences due to the ramdomness. CENT$_{PC1}$ and CENT$_{PLS}$ stand for the centroid classifier (\ref{centroid-asymp}), where the function $\psi$ is estimated via principal components or PLS components, respectively. NP refers to the classifier based in the non-parametric functional regression method proposed by \cite{fer06} and CENT$_{PCp}$ denotes the usual centroid classifier applied to the multivariate principal component projections. The outputs correspond to the average (over 200 runs) percentages of misclassification obtained for each method, sample size and data set. The values in parentheses correspond to the standard deviation of these errors.

\begin{table}
	\begin{center}\footnotesize
		\caption{\footnotesize Misclassification percentages (and standard deviations) for the classification methods considered in Table 2 of \cite{del12} and the new RK-C method} \medskip \label{table:real2}
		\begin{tabular}{llccccc}
			\toprule
			\small Data &\small $n$ & \multicolumn{5}{c}{\small Classification rules}  \\ \cmidrule{3-7}
			&  & CENT$_{PC1}$ & CENT$_{PLS}$ & NP & CENT$_{PCp}$ & \textbf{RK-C} \\ 
			\midrule
			Wheat & 30 & 0.89 (2.49) & 0.46 (1.24) & 0.49 (1.29) & 15.0 (1.25) & \textbf{0.25 (1.58) }\\ 
			& 50 & 0.22 (1.09) & 0.06 (0.63) & 0.01 (0.14) & 14.4 (5.52) & \textbf{0.02 (0.28)} \\ 
			Phoneme & 30 & 22.5 (3.59) & 24.2 (5.37) & 24.4 (5.31) & 23.7 (2.37) & \textbf{22.5 (3.70) }\\ 
			& 50 & 20.8 (2.08) & 21.5 (3.02) & 21.9 (2.91) & 23.4 (1.80) & \textbf{21.5 (2.36)} \\ 
			& 100 & 20.0 (1.09) & 20.1 (1.12) & 20.1 (1.37) & 23.4 (1.36) & \textbf{20.1 (1.25)} \\ 
			\bottomrule
		\end{tabular} 
	\end{center}
\end{table}
\normalsize

The results show that the RK-C classifier is clearly competitive against the remaining methods. In addition, there is perhaps some interpretability advantage in the use of RK-C, as this method is based in dimension reduction via variable selection so that the "reduced data" are directly interpretable in terms of the original variables. Let us finally point out that the variable selection process is quite efficient: in the wheat example, near perfect classification is achieved using just one variable; in the much harder phoneme example, the average number of selected variables is three.

\section{Conclusions}\label{sec:conclusions}

We have proposed a RKHS-based method for both variable selection and binary classification. It is fully theoretically motivated in terms of the RKHS space associated with the underlying model. 
We next summarize our study of the RK methods in the following conclusions.

\begin{itemize}
	\item[a)] The identification of the RKHS associated with a supervised classification problem represents several important theoretical and practical advantages. Apart from providing explicit expressions of the optimal Bayes rule (via the corresponding Radon-Nikodym derivatives), the RKHS approach provides a theoretical explanation for the   near perfect classification phenomenon in terms of the mutual singularity of the involved measures. 
	\item[b)] Perhaps more importantly, the RKHS approach provides  a theoretical scenario to motivate the use of variable selection. Under the RKHS framework, the family of models fulfilling a finite RKHS expansion for $m$ of type $m(\cdot)=\sum_{i=1}^dK(\cdot,t_i)$ is dense in the whole class of considered models. Note also that, even if a finite expansion is not exactly fulfilled, the method has a clear interpretation (see the comments after Proposition \ref{prop:explanation}) as it looks for the ``best'' choice of $(t_1,\ldots,t_d)$ under this approximated model. The point is that, in any case, the method is always motivated in population terms.  
	\item[c)] The RKHS-based variable selection and classification procedures are quite accurate and computationally inexpensive with important  advantages in terms of simplicity and interpretability.
	The simulation outputs show that RK-VS procedure is especially successful as a variable selection method. As a classifier RK-C is still competitive and especially good when the underlying assumptions are fulfilled. 
	\item[d)] The empirical results show also a remarkable robustness of the RK methodology against departures from the assumptions on which it is based. 
\end{itemize}

\noindent \textbf{Acknowledgements}. Research partially supported by Spanish grant MTM2013-44045-P.

\noindent \textbf{Supplementary materials}. The file \textit{Supplementary.pdf} includes all the proofs  as well as some additional details, tables and figures on the empirical results. The file \textit{Outputs.xlsx} gives the complete simulation outputs.

\spacingset{1.1}

\newpage

\begin{center}
	\large \bf  Supplementary material for the paper ``On the use of reproducing kernel Hilbert spaces in functional classification''
\end{center}
\normalsize


\def\theequation{S\arabic{equation}}
\def\thesection{S\arabic{section}}
\def\thefigure{S\arabic{figure}}
\def\thetable{S\arabic{table}}

\setcounter{equation}{0} 
\setcounter{section}{0} 
\setcounter{figure}{0} 
\setcounter{table}{0} 
\section{Proofs}\label{proofs}

\begin{proof}[Proof of Theorem 2]
	
	Equation (4) follows straightforwardly from the combination of (1) and (2). To prove the expression for the Bayes error notice that
	$\innerk{X-m_0}{m}$ lies in $\bar{\mathcal L}(X-m_0)$ and therefore the random variable $\eta^*(X)$ is Gaussian both under $Y=1$ and $Y=0$. Furthermore, Equations (6.19) and (6.20) in \cite{par61} yield
	\begin{align*}
	\mathbb{E}(\eta^*(X) | Y=0)&= -\normk{m}^2/2-\log\left(\frac{1-p}{p}\right),\\
	\mathbb{E}(\eta^*(X) | Y=1) &=  \normk{m}^2/2-\log\left(\frac{1-p}{p}\right),\\
	\mbox{Var}(\eta^*(X) | Y=0) &= \mbox{Var}(\eta^*(X) | Y=1) = \normk{m}^2.
	\end{align*}
	The result follows using these values to standardize the variable $\eta^*(X)$ in 
	$L^* = (1-p)\mathbb{P}(\eta^*(X)>0 | Y=0) + p \mathbb{P}(\eta^*(X)<0 | Y=1)$.
\end{proof}
\begin{proof}[Proof of Theorem 4]
	
	Observe that, if $\theta_j>0$ for all $j\geq 1$,
	\[
	m_1 = \sum_{j=1}^\infty \mu_j\phi_j = \sum_{j=1}^\infty \frac{\mu_j}{\sqrt{\theta_j}} \sqrt{\theta_j}\phi_j,
	\]
	where $\{\sqrt{\theta_j}\phi_j:\, \theta_j>0\}$ is an orthonormal basis of $\mathcal{H}(K)$ [see, e.g., Theorem 4.12, p. 61 in \cite{cucker07}]. Then, by Parseval's formula, $m_1\in\mathcal{H}(K)$ if and only if $\normk{m_1}^2=\sum_{j=1}^\infty \theta^{-1}_j\mu_j^2 < \infty$. As a consequence, we have the desired equivalence:
	\[
	P_1\sim P_0 \Leftrightarrow m_1\in \mathcal{H}(K) \Leftrightarrow \normk{m_1}<\infty \Leftrightarrow \sum_{j=1}^\infty \theta^{-1}_j\mu_j^2<\infty.
	\]
	Moreover,
	\[
	\mbox{err}_0 = 1-\Phi\left(\frac{1}{2}(\sum_{j=1}^\infty\theta^{-1}_j \mu_j^2)^{1/2}\right)=1-\Phi\left(\frac{1}{2}\normk{m_1}\right),
	\]
	what gives the coordinate-free expression of the Bayes error.
	
	Now, if we further assume (as in \cite{del12np}) that $\psi\in L^2$, the optimal classifier proposed by these authors (5) is equivalent to $T^0(X)=1$ if and only if
	\begin{equation}
	\label{eq:proof6.1}
	\innerl{m_1}{\psi}^2 - 2\innerl{m_1}{\psi}\innerl{X}{\psi} < 0.
	\end{equation}
	Since $m_1=\sum_{j=1}^\infty \mu_j\phi_j$, with $m_1\neq 0$, and $\psi=\sum_{j=1}^\infty \theta_j^{-1}\mu_j\phi_j$, we have  $\innerl{m_1}{\psi}=\sum_{j=1}^\infty\theta^{-1}_j \mu_j^2=\normk{m_1}^2\neq 0$. Therefore,  (\ref{eq:proof6.1}) holds if and only if 
	\[
	\innerl{X}{\psi} - \frac{\normk{m_1}^2}{2} > 0.
	\] 
	To end the proof it is enough to show $\innerk{X}{m_1}=\innerl{X}{\psi}$. The linearity of $\innerk{X}{\cdot}$ and the fact that $\theta_j$ and $\phi_j$ are respectively eigenvalues and eigenfunctions of the integral operator with kernel $K$ imply
	\[
	\innerk{X}{m_1}= \sum_{j=1}^\infty \theta_j^{-1}\mu_j\innerk{X}{\theta_j\phi_j} =
	\sum_{j=1}^\infty \theta_j^{-1}\mu_j \int_0^T \innerk{X}{K(\cdot,u)}\phi_j(u)du.
	\] 
	Now, from Equation (6.18) in \cite{par61},
	\[
	\int_0^T \innerk{X}{K(\cdot,u)}\phi_j(u)du = \int_0^T X(u)\phi_j(u)du = \innerl{X}{\phi_j}.
	\]
	Finally, combining the two last displayed equations,
	\[
	\innerk{X}{m_1} = \sum_{j=1}^\infty \theta_j^{-1}\mu_j\innerl{X}{\phi_j}=
	\innerl{X}{\sum_{j=1}^\infty \theta_j^{-1}\mu_j\phi_j}=\innerl{X}{\psi}.
	\]
\end{proof}

\

\begin{proof}[Proof of Theorem 5]
	Let $X=\sum_{i=1}^\infty Z_i\phi_i$, the Karhunen-Loève expansion of $X$, with the $Z_i$ uncorrelated. 
	For a given trajectory $x=\sum_{i=1}^\infty z_i\phi_i$. Define
	$x^n=\sum_{i=1}^n z_i\phi_i$,
	This is a trajectory drawn from the process 
	$X^n=\sum_{i=1}^n Z_i\phi_i$,
	whose distribution under $P_i$ is denoted by $P_{in}$ (for $i=0,1$, the covariance function is $K_n(s,t)=\sum_{i=1}^n {\mathbb E}(Z_i^2)\phi_i(s)\phi_i(t)$ and the mean function under $P_{1n}$ is 
	$$
	m_n(t)=\sum_{i=1}^n {\mathbb E}(Z_i)\phi_i(t),
	$$
	Note that, under $P_0$, ${\mathbb E}(Z_j)=0$, so that the mean function is 0. 
	From Karhunen-Loève Theorem (see \cite {ash75}, p. 38) $m_n(t)\to m(t)$ for all $t$ (in fact this results holds uniformly in $t$). 
	
	Note also that $m_n\in {\mathcal H}(K)$. Again this follows from the fact that  $\{\sqrt{\theta_i}\phi_i:\, \theta_i>0\}$ is an orthonormal basis of $\mathcal{H}(K)$ [see, e.g., Theorem 4.12, p. 61 in \cite{cucker07}].
	
	We now prove that we must necessarily have $\lim_n\Vert m_n\Vert_K=\infty$. Indeed, if we had $\lim_n\Vert m_n\Vert_K<\infty$ for some subsequence
	of $\{m_n\}$ (denoted again $\{m_n\}$) we would have that  such $\{m_n\}$ would be a Cauchy sequence in 
	${\mathcal H}(K)$, since for $q>p$, $\Vert m_p-m_q\Vert_K\leq |\Vert m_q\Vert_K-\Vert m_p\Vert_K|$. This, together with the pointwise convergence $m_n(t)\to m(t)$ leads, from Moore-Aronszajn Theorem (see \cite{ber04}, p. 19) to $m\in {\mathcal H}(K)$. But, from Parzen's Theorem 1, this would entail $P_1<<P_0$, in contradiction with $P_1\perp P_0$. We thus conclude $\Vert m_n\Vert_K\to\infty$.

	Then, given $\epsilon>0$, choose $n$ 
	such that 
	{\small\begin{eqnarray}
		&&(1-p) \Phi\left(-\frac{\normk{m_n}}{2}-\frac{1}{\normk{m_n}}\log\left(\frac{1-p}{p} \right) \right)\nonumber \\
		&&+ p \Phi\left(-\frac{\normk{m_n}}{2}+\frac{1}{\normk{m_n}}\log\left(\frac{1-p}{p} \right) \right)<\epsilon,\label{error}
		\end{eqnarray}}

	Now, consider the problem $X^n\sim P_{1n}$ vs $X^n\sim P_{0n}$
	Note that $X^n\sim P_{in}$ if and only if  $X\sim P_i$, for $i=0,1$. Since $m_n\in {\mathcal H}(K_n)$, we have
	$P_{0n}\sim P_{1n}$ (using again Parzen's Theorem 1). 
	
	Now, according to Theorem 2 (on the expression of the optimal rules in the absolutely continuous case under homoscedasticity), the optimal rule is
	$g_n(X)=\mathbb{I}_{\{\eta_n(X)>0\}}$, where
	{\small \begin{equation}
		\eta_n(x) = 
		\innerk{x}{m_n} - \frac{1}{2}\normk{m_n}^2 - \log\left(\frac{1-p}{p} \right),
		\end{equation}}
	whose probability of error, is exactly the expression on the left-hand side of (\ref{error}).
	So this probability can be made arbitrarily small.
\end{proof}

\begin{proof}[Proof of Theorem 6]
	
	For the sake of conciseness, denote $\tau:= (t_1,\ldots,t_d)$, a generic element of $\Theta$, $\hat \tau:= (\hat t_1,\ldots,\hat t_d)$, and $\tau^*:= (t^*_1,\ldots,t^*_d)$. We will also use the following notation: for $j=0,1$,
	\[
	\tilde \psi_j(\tau) := \frac{\big(2(m_{j,\tau}-\hat \mu_\tau)^\top \hat K^{-1}_\tau \hat{m}_\tau\big)^2}{\hat{m}_\tau^\top \hat{K}^{-1}_\tau K_\tau \hat{K}^{-1}_\tau\hat{m}_\tau},
	\]
	where $m_{j,\tau}:=(m_j(t_1),\ldots, m_j(t_d))^\top$ and $\hat{\mu}_\tau = (\hat{m}_{0,\tau}+\hat{m}_{1,\tau})/2$. With this notation it is not difficult to show that $L^*=1-\Phi(\psi(\tau^*)^{1/2}/2)$, and  
	\[
	L_n=1-\frac{1}{2}\Phi\left(\frac{\tilde\psi_0(\hat\tau)^{1/2}}{2}\right)-\frac{1}{2}\Phi\left(\frac{\tilde\psi_1(\hat\tau)^{1/2}}{2}\right),
	\]
	where $\Phi$ is the cumulative distribution function of the standard Gaussian distribution (to obtain these formulas we have used the arguments in  \cite{mardia} 
	p.  321, for $L^*$, and \cite{FanFan}, p. 2609, for $L_n$).  Since $\Phi$ is continuous, the desired conclusion 
	will readily follow if we prove 
	$\tilde \psi_j(\hat{\tau})\to \psi(\tau^*)$ as $n\to\infty$, a.s., for $j=0,1$.

	Since $\mathbb{E}\|\epsilon_j\|_\infty<\infty$, for $j=0,1$, Mourier's Strong Law of Large Numbers (SLLN) for random elements taking values in Banach spaces (see e.g.  \cite{lah79}, p. 452)  implies
	\begin{equation}
	\label{eq:unifm}
	\sup_{\tau\in \Theta} \|\hat{m}_{\tau}-m_{\tau}\|\to 0,\ \ \mbox{as}\ n\to\infty,\ \ \ \mbox{a.s.}
	\end{equation}
	Since  $\mathbb{E}\|\epsilon_j^2\|_\infty<\infty$ for $j=0,1$, Mourier's SLLN also implies that the entries of $\hat{K}_{\tau}$  converge uniformly to those of $K_{\tau}$, that is for $i,j=1,\ldots,d$,
	\begin{equation}
	\label{eq:unifKij}
	\sup_{\tau\in \Theta} |\hat{K}_{\tau}(i,j)-K_{\tau}(i,j) |\to 0,\ \ \mbox{as}\ n\to\infty,\ \ \ \mbox{a.s.}
	\end{equation}
	Observe that
	\[
	\hat{K}^{-1}_{\tau} = \frac{\mbox{adj}(\hat{K}_{\tau})}{\mbox{det}(\hat{K}_{\tau})},
	\]
	where $\mbox{adj}(K)$ and $\mbox{det}(K)$ denote the adjugate and the determinant of a matrix $K$, respectively. By (\ref{eq:unifKij}), the entries of $\mbox{adj}(\hat{K}_{\tau})$  converge uniformly to those of $\mbox{adj}(K_\tau)$, and   
	$\mbox{det}(\hat{K}_{\tau})$ converges uniformly to $\mbox{det}(K_{\tau})$. Moreover,  $\inf_{\tau\in \Theta} \mbox{det}(K_\tau) > 0$ because $\mbox{det}(K_\tau)$  is continuous in $\tau$  and, by assumption, $\mbox{det}(K_\tau)>0$, for all $\tau\in \Theta$, where $\Theta$ is a compact set. As a consequence of all these observations,
	\begin{equation}
	\label{eq:unifKinvij}
	\sup_{\tau\in \Theta} |\hat{K}^{-1}_{\tau}(i,j)-K^{-1}_{\tau}(i,j) |\to 0,\ \ \mbox{as}\ n\to\infty,\ \ \ \mbox{a.s.}
	\end{equation}
	By (\ref{eq:unifm}) and (\ref{eq:unifKinvij}), it also holds
	\[
	\sup_{\tau\in \Theta} \| \hat{K}^{-1}_\tau \hat{m}_\tau- K^{-1}_\tau m_\tau\|\to 0,\ \ \mbox{as}\ n\to\infty,\ \ \ \mbox{a.s.}
	\]
	From this convergence, together with (\ref{eq:unifm}), we deduce
	\begin{equation}
	\sup_{\tau\in \Theta} |\hat \psi(\tau) -\psi(\tau)|\to 0,\ \ \mbox{as}\ \ n\to\infty, \ \ \mbox{a.s.} 
	\label{eq.unif}
	\end{equation}
	and
	\begin{equation}
	\sup_{\tau\in \Theta} |\tilde \psi_j(\tau) -\psi(\tau)|\to 0,\ \ \mbox{as}\ \ n\to\infty, \ \ \mbox{a.s.} \ \ \ j=0,1.
	\label{eq.uniftilde}
	\end{equation}
	Due to (\ref{eq.unif}), with probability one, given $\epsilon>0$ there exists $N$ such that for $n\geq N$ it holds
	$
	\hat{\psi}(\tau) - \epsilon \leq \psi(\tau) \leq \hat{\psi}(\tau) + \epsilon,
	$ for all $\tau\in \Theta$.
	Taking the maximum  in these inequalities we get
	$
	\hat{\psi}(\hat{\tau}) - \epsilon \leq \psi(\tau^*) \leq \hat{\psi}(\hat{\tau}) + \epsilon$. That is, we have 
	\begin{equation}
	\hat{\psi}(\hat \tau)\to \psi(\tau^*),\ \ \mbox{as } n\to \infty\ \ \ \mbox{a.s}.\label{psi}
	\end{equation}
	Finally, note that for $j=0,1$,
	\[
	|\tilde \psi_j(\hat{\tau}) - \psi(\tau^*)|\leq |\tilde \psi_j(\hat{\tau}) -\psi(\hat{\tau})| + |\psi(\hat{\tau})-\hat \psi(\hat{\tau})| + |\hat\psi(\hat{\tau}) - \psi(\tau^*)|.
	\]
	Then, from (\ref{eq.unif}), (\ref{eq.uniftilde}) and (\ref{psi}) we get $\tilde \psi_j(\hat{\tau})\to \psi(\tau^*)$ as $n\to\infty$, a.s. for $j=0,1$, as desired.
\end{proof}

\section{Models used in the simulation study}

The general structure is similar to that of the simulation studies in \cite{ber15b} and \cite{ber15a} which are devoted to the assessment of variable selection methods in the functional classification setting.   Here we consider the 94 models for which the mean functions $m_0$ and $m_1$ are different.
The optimal classification rule in each case depends only on a finite number of variables. 
Models differ in complexity and number of relevant variables. They are defined giving either:
\begin{itemize}
	\item[(E1)] A pair of distributions for $X|Y=0$ and $X|Y=1$ (corresponding to $P_0$ and $P_1$, respectively) as well as the prior probability $p=\mathbb{P}(Y=1)$; in all cases, we take $p={\mathbb P}(Y=1)=1/2$.
	\item[(E2)] The marginal distribution of $X$ plus the conditional distribution $\eta(x)={\mathbb P}(Y=1|X=x)$. 
\end{itemize}
All the 94 considered models belong to one of the following classes:
\paragraph{Gaussian models:} they are denoted by $G$. Gaussian models are generated according to the general pattern (E1). In all cases the distributions of $X(t)|Y=i$ are chosen among one of the Gaussian distributions described below.

\paragraph{Logistic models:} they are defined through the general pattern (E2). The process $X=X(t)$ follows one of the above mentioned distributions and $Y\sim\mbox{Binom}(1,\eta(X))$ with  
$$\eta(x)=\frac{1}{1+e^{-\Psi(x(t_1),\cdots,x(t_d))}},$$
a function of the relevant variables $x(t_1),\cdots,x(t_d)$. The 15 versions and the few variants of this model considered are identified with the general label $L$. They correspond to different choices for the link function $\Psi$  (both linear and nonlinear) and for the distribution of $X$.  

\paragraph{Mixtures:} they are obtained by combining (via mixtures) the above mentioned Gaussian distributions assumed for $X|Y=0$ and $X|Y=1$ in several ways. These models are denoted by $M$ in the output tables. 

\

The processes involved are chosen among the following: first, the \bf standard Brownian Motion\rm, $B$. Second,
$BT$ denotes a \bf Brownian Motion with a trend \rm $m(t)$, i.e., $BT(t)$ $=B(t)+m(t)$; we have considered several choices for $m(t)$, a linear trend, $m(t)=ct$, a linear trend with random slope, i.e., $m(t)=\theta t$, where $\theta$ is a Gaussian r.v., and  different members of two  
parametric families: the \it peak\/ \rm functions
$\Phi_{m,k}$ and the \it hillside\/ \rm functions, defined by 
$$
\Phi_{m,k}=\int_0^t\varphi_{m,k}(s)ds \hspace{15pt} , \hspace{15pt} \mbox{hillside}_{t_0 ,b}(t)=b(t-t_0){\mathbb I}_{[t_0,\infty)},
$$
where, $\varphi_{m,k}(t)=\sqrt{2^{m-1}}\left[\mathbb{I}_{\left(\frac{2k-2}{2^m},\frac{2k-1}{2^m}\right)}-\mathbb{I}_{\left(\frac{2k-1}{2^m},\frac{2k}{2^m}\right)}\right]$ for $m\in\mathbb{N}$, 
$1\leq k\leq 2^{m-1}$.
Third, the \bf Brownian Bridge\rm: $BB(t)=B(t)-tB(1)$. Our fourth class of Gaussian processes is the \bf Ornstein–Uhlenbeck process\rm, with  zero mean ($OU$) or different mean functions $m(t)$ ($OUt$). Finally some ``smooth'' processes have been also included. They are obtained by convolving Brownian trajectories with Gaussian kernels. We have considered two levels of smoothing denoted by sB and ssB; in the  list of models below those labeled ssB are smoother than those with label sB.

In the following list of models, $P_i$ denotes the distribution of $X|Y=i$ and \textit{variables} is the set of relevant variables in each Gaussian or Mixture case. We call them ``relevant'' in the sense that the optimal classification rule depends only on these variables. In the list below the variables written in boldface are ``especially relevant'' in terms of their relative discriminating capacity.

All considered sample data are discretized in 100 equispaced points $X_1,\ldots, X_100$ in the interval [0,1]. To avoid degeneracies we have excluded the point 0 and the point 1 in the Brownian Bridge type models. 

\

\noindent 1. \sc Gaussian models considered\rm:

\small
\begin{multicols}{2}
	\renewcommand{\labelenumi}{\scriptsize\arabic{enumi}.}
	\begin{enumerate}
		%
		%
		%

		\item \textbf{G2} : $	\left \{ 
		\begin{matrix}
		P_0:&B(t)+ t&\\
		P_1:&B(t)&
		\end{matrix}\right. $
		\item[] $variables=\{X_{100}\}$.
		

		\item \textbf{G2b} : $	\left \{ 
		\begin{matrix}
		P_0:&B(t)+ 3t&\\
		P_1:&B(t)&
		\end{matrix}\right. $
		\item[] $variables=\{X_{100}\}$.
		

		
		
		\item \textbf{G4} : $		\left \{ 
		\begin{matrix}
		P_0:&B(t)+hillside_{0.5,4}(t)&\\
		P_1:&B(t)&
		\end{matrix}\right.$
		\item[] $variables=\{X_{47},$\textbf{\textit{X}}$_{100}\}$.

		\item \textbf{G5} : $		\left \{ 
		\begin{matrix}
		P_0:&B(t)+3\Phi_{1,1}(t)&\\
		P_1:&B(t)&
		\end{matrix}\right.$
		\item[] $variables=\{X_1,$\textbf{\textit{X}}$_{48},X_{100}\}$.
		
		\
		
		\item \textbf{G6} : $		\left \{ 
		\begin{matrix}
		P_0:&B(t)+5\Phi_{2,2}(t)&\\
		P_1:&B(t)&
		\end{matrix}\right.$
		\item[] $variables=\{X_{48},$\textbf{\textit{X}}$_{75},X_{100}\}$.

		\item \textbf{G7} : $		\left \{ 
		\begin{matrix}
		P_0:&B(t)+5\Phi_{3,2}(t)+5\Phi_{3,4}(t)&\\
		P_1:&B(t)&
		\end{matrix}\right.$
		\item[] $variables=\{X_{22},$\textbf{\textit{X}}$_{35},X_{49},X_{74},$\textbf{\textit{X}}$_{88},X_{100}\}$.
		
		\
		
		\item \textbf{G8} :  $		\left \{ 
		\begin{matrix}
		P_0:&B(t)+3\Phi_{2,1.25}(t)+3\Phi_{2,2}(t)&\\
		P_1:&B(t)&
		\end{matrix}\right.$
		\item[] $variables=\{X_{9},$\textbf{\textit{X}}$_{35},X_{48},X_{62},$\textbf{\textit{X}}$_{75},X_{100}\}$.
		
	\end{enumerate}
	
\end{multicols}
\rm

\noindent 2. \sc logistic-type models under study\rm: they are all defined according method (E2) (see Sec. 6.1 in the main paper). The process $X=X(t)$ follows one of the distributions mentioned above and $Y=\mbox{Binom}(1,\eta(X))$ with 
$\eta(x)=(1+e^{-\psi(x(t_1),\cdots,x(t_k))})^{-1}$,
a function of the relevant variables $x(t_1),\cdots,x(t_k)$. 

\

\textbf{L1:} $\psi(X)=10X_{65}$.

\

\textbf{L2:} $\psi(X)=10X_{30} + 10X_{70}$.

\

\textbf{L3:} $\psi(X)=10X_{30} - 10X_{70}$.

\

\textbf{L4:} $\psi(X)=20X_{30} + 50X_{50} 20X_{80}$.

\

\textbf{L5:} $\psi(X)=20X_{30} - 50X_{50}+ 20X_{80}$.

\

\textbf{L6:} $\psi(X)=10X_{10} + 30X_{40} + 10X_{72}+ 10X_{80} +20X_{95}$.

\

\textbf{L7:} $\psi(X)= \sum_{i=1}^{10} 10X_{10i}$.

\

\textbf{L8:} $\psi(X)= 20X_{30}^2 + 10X_{50}^4 + 50X_{80}^3$.

\

\textbf{L9:} $\psi(X)= 10X_{10} + 10|X_{50}| + 0X_{30}^2 X_{85}$.

\

\textbf{L10:} $\psi(X)= 20X_{33} + 20|X_{68}|$.

\

\textbf{L11:} $\psi(X)= \frac{20}{X_{35}}+\frac{30}{X_{77}}$.

\

\textbf{L12:} $\psi(X)= \log{X_{35} + \log{X_{77}}}$.

\

\textbf{L13:} $\psi(X)= 40X_{20}+30X_{28}+20X_{62}+10X_{67}$.

\

\textbf{L14:} $\psi(X)= 40X_{20}+30X_{28}-20X_{62}-10X_{67}$.

\

\textbf{L15:} $\psi(X)= 40X_{20}-30X_{28}+20X_{62}-10X_{67}$.

\

Some variations of these models have been also considered:

\

\textbf{L3b:} $\psi(X)=30X_{30} - 20X_{70}$.

\

\textbf{L4b:} $\psi(X)=30X_{30} + 20X_{50}+ 10X_{80}$.

\

\textbf{L5b:} $\psi(X)=10X_{30} - 10X_{50}+ 10X_{80}$.

\

\textbf{L6b:} $\psi(X)=20X_{10} + 20X_{40} + 20X_{72}+ 20X_{80} +20X_{95}$.

\

\textbf{L8b:} $\psi(X)=10X_{30}^2 + 10X_{50}^4+ 10X_{80}^3$.

\

\noindent 3. \sc Mixture-type models\rm: they are obtained by combining (via mixtures) in several ways the above mentioned Gaussian distributions assumed for $X|Y=0$ and $X|Y=1$. These models are denoted M1, ..., M10 in the output tables. 

\scriptsize
\begin{multicols}{2}
	\renewcommand{\labelenumi}{\scriptsize\arabic{enumi}.}
	
	\begin{enumerate}
		
		
		\item \textbf{M2} : $		\left \{ 
		\begin{matrix}
		P_0:&\left \{ 
		\begin{matrix}
		&B(t)+3\Phi_{2,2}(t),& 1/2\\
		&B(t)+5\Phi_{3,2}(t),& 1/2
		\end{matrix}\right.\\
		&&\\
		P_1:&B(t)& 
		\end{matrix}\right.$
		\item[] $variables=\{X_{22},$\textbf{\textit{X}}$_{35},X_{48},$\textbf{\textit{X}}$_{75},X_{100}\}$.
		
		\
		
		\item \textbf{M3} : $		\left \{ 
		\begin{matrix}
		P_0:&\left \{ 
		\begin{matrix}
		&B(t)+3\Phi_{2,2}(t),& 1/10\\
		&B(t)+5\Phi_{3,2}(t),& 9/10
		\end{matrix}\right.\\
		&&\\
		P_1:&B(t)& 
		\end{matrix}\right.$
		\item[] $variables=\{X_{22},$\textbf{\textit{X}}$_{35},X_{48},$\textbf{\textit{X}}$_{75},X_{100}\}$.

		\
		
		\item \textbf{M4}: $		\left \{ 
		\begin{matrix}
		P_0:&\left \{ 
		\begin{matrix}
		&B(t)+3\Phi_{2,2}(t),& 1/2\\
		&B(t)+5\Phi_{3,3}(t),& 1/2
		\end{matrix}\right.\\
		&&\\
		P_1:&B(t)& 
		\end{matrix}\right.$
		\item[] $variables=\{X_{48},$\textbf{\textit{X}}$_{62},$\textbf{\textit{X}}$_{75},X_{100}\}$.

		\
		
		\item \textbf{M5} :$		\left \{ 
		\begin{matrix}
		P_0:&\left \{ 
		\begin{matrix}
		&B(t)+3\Phi_{2,1}(t)&, 1/3\\
		&B(t)+3\Phi_{2,2}(t),& 1/3\\
		&B(t)+5\Phi_{3,2}(t),& 1/3
		\end{matrix}\right.\\
		&&\\
		P_1:&B(t)& 
		\end{matrix}\right.$
		\item[] $variables=\{X_{1},$\textbf{\textit{X}}$_{22},$\textbf{\textit{X}}$_{35},X_{48},$\textbf{\textit{X}}$_{75},X_{100}\}$.
		
		\columnbreak
		
		\item \textbf{M6} : $		\left \{ 
		\begin{matrix}
		P_0:&\left \{ 
		\begin{matrix}
		&B(t)+3\Phi_{2,1}(t)&, 1/2\\
		&B(t)+3t&, 1/2
		\end{matrix}\right.\\
		&&\\
		P_1:&B(t)& 
		\end{matrix}\right.$
		\item[] $variables=\{X_{1},$\textbf{\textit{X}}$_{22},X_{49},$\textbf{\textit{X}}$_{100}\}$.
		
		\

		\item \textbf{M7} : $		\left \{ 
		\begin{matrix}
		P_0:&\left \{ 
		\begin{matrix}
		&B(t)+3\Phi_{1,1}(t)&, 1/2\\
		&BB(t)&, 1/2
		\end{matrix}\right.\\
		&&\\
		P_1:&B(t)& 
		\end{matrix}\right.$
		\item[] $variables=\{X_{1},$\textbf{\textit{X}}$_{48},$\textbf{\textit{X}}$_{100}\}$.
		
		\

		\item \textbf{M8} : $		\left \{ 
		\begin{matrix}
		P_0:&\left \{ 
		\begin{matrix}
		&B(t)+\theta t,\hspace{5pt} \theta \sim N(0,5)&, 1/2\\
		&B(t)+hillside_{0.5,5}(t)&, 1/2
		\end{matrix}\right.\\
		&&\\
		P_1:&B(t)& 
		\end{matrix}\right.$
		\item[] $variables=\{X_{47},$\textbf{\textit{X}}$_{100}\}$.


		\

		\item \textbf{M10} : $		\left \{ 
		\begin{matrix}
		P_0:&\left \{ 
		\begin{matrix}
		&B(t)+3\Phi_{1,1}(t)&, 1/3\\
		&B(t)-3t&, 1/3\\
		&BB(t)&, 1/3
		\end{matrix}\right.\\
		&&\\
		P_1:&B(t)& 
		\end{matrix}\right.$
		\item[] $variables=\{X_{1},$\textbf{\textit{X}}$_{48},$\textbf{\textit{X}}$_{100}\}$.
		

	\end{enumerate}
	\rm
\end{multicols}

\

\normalsize
Finally, we consider here those models for which the mean functions $m_0$ and $m_1$ are different (otherwise any linear method is blind to discriminate between $P_0$ and $P_1$). The full list of models involved is  as follows:

\renewcommand{\labelenumi}{\scriptsize\arabic{enumi}.}
\small
\begin{multicols}{4}
	\columnseprule 0.25pt
	\begin{enumerate}
		\item L1 OU
		\item L1 OUt
		\item L1 B
		\item L1 sB
		\item L1 ssB
		\item L2 OU
		\item L2 OUt
		\item L2 B
		\item L2 sB
		\item L2 ssB
		\item L3 OU
		\item L3b OU
		\item L3 OUt
		\item L3b OUt
		\item L3 B
		\item L3b B
		\item L3 sB
		\item L3 ssB
		\item L4 OU
		\item L4b OU
		\item L4 OUt
		\item L4b OUt
		\item L4 B
		\item L4 sB
		\item L4 ssB
		\item L5 OU
		\item L5b OU
		\item L5 OUt
		\item L5 B
		\item L5 sB
		\item L5 ssB
		\item L6 OU
		\item L6b OU
		\item L6 OUt
		\item L6b OUt
		\item L6 B
		\item L6 sB
		\item L6 ssB
		\item L7 OU
		\item L7b OU
		\item L7 OUt
		\item L7b OUt
		\item L7 B
		\item L7 sB
		\item L7 ssB
		\item L8 B
		\item L8 sB
		\item L8 ssB
		\item L8b OU
		\item L9 B
		\item L9 sB
		\item L9 ssB
		\item L10 OU
		\item L10 B
		\item L10 sB
		\item L10 ssB
		\item L11 OU
		\item L11 OUt
		\item L11 B
		\item L11 sB
		\item L11 ssB
		\item L12 OU
		\item L12 OUt
		\item L12 B
		\item L12 sB
		\item L12 ssB
		\item L13 OU
		\item L13 OUt
		\item L13 B
		\item L13 sB
		\item L13 ssB
		\item L14 OU
		\item L14 OUt
		\item L14 B
		\item L14 sB
		\item L15 OU
		\item L15 OUt
		\item L15 B
		\item L15 sB
		\item G2
		\item G2b
		\item G4
		\item G5
		\item G6
		\item G7
		\item G8
		\item M2
		\item M3
		\item M4
		\item M5
		\item M6
		\item M7
		\item M8
		\item M10
	\end{enumerate}
\end{multicols}

\section{Computational details}\label{sec:comp}

All considered methodologies have been implemented in MATLAB. The code is available upon request. Some details:
\begin{itemize}
	\item We have followed the implementation of the the minimum Redundancy Maximum Relevance algorithm given in \cite{ber15a}. This version allows us to introduce different association measures.
	
	\item We have implemented the original iterative PLS algorithm that can be found, e.g. in
	\cite{del12pls}. 
	
	\item Maxima-hunting and the distance correlation measure have been computed as described in \cite{ber15b}.
	
	\item Our $k$-NN implementation is built around the MATLAB function \textit{pdist2} and allows for the use of different distances; we have employed the usual Euclidean distance. Also, the computation for different numbers of neighbours can be simultaneously made with no additional cost.
	
	\item Our LDA is a faster implementation of the MATLAB function \textit{classify}.
	
	\item The linear SVM has been performed with the MATLAB version of the LIBLINEAR library (see \cite{fan08svm}) using the parameters \textit{bias} and \textit{solver type} 2. It obtains (with our data) very similar results to those of the default \textit{solver type} 1, but faster. LIBLINEAR is much faster than the more popular LIBSVM library when using linear kernels.
	
	\item The  cost parameter $C$ of the linear SVM classifier, the number $k$ of nearest neighbours in the $k$-NN rule, the smoothing parameter $h$ in MHR and the number of selected variables are chosen by standard validation procedures explained in Section 6. 
\end{itemize}

\section{Additional results}

\begin{figure}
	\centering
	\begin{minipage}{0.45\textwidth}
		\centering 		 
		\includegraphics[width=1\linewidth]{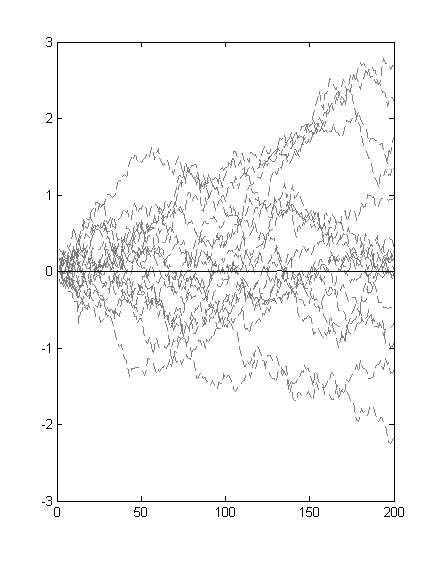}
	\end{minipage}\hfill
	\begin{minipage}{0.45\textwidth}
		\centering
		\includegraphics[width=1\linewidth]{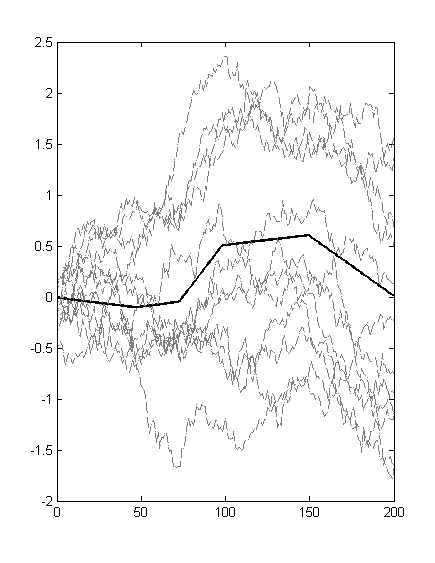} 		
	\end{minipage}
	\caption{\footnotesize Some trajectories from the toy example $B(t)$ (left) vs $B(t)+\Phi_{1,1}(t)-\Phi_{2,1}(t) +\Phi_{2,2}(t) - \Phi_{3,2}(t)$ (right). Thick solid lines correspond  to the  mean functions.}\label{fig:trayectories_ej}
\end{figure}

\begin{figure}
	\centering
	\includegraphics[width=1\linewidth]{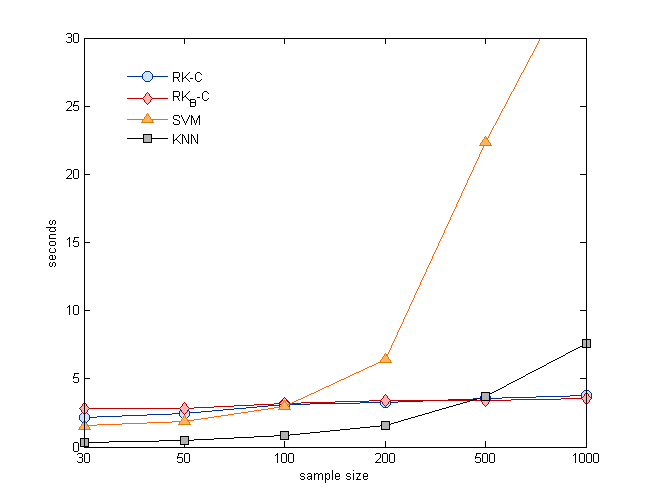} 	
	\caption{\footnotesize Time cost (in seconds) of 100 runs of the experiment for each method and different sample sizes with $d=200$.}\label{fig:cost}
\end{figure}

In this section we include some supplementary outputs and graphs of practical interest as well as more detailed information about the simulation results:

\begin{itemize}
	\item Some trajectories of the toy example in Section 5.5 are displayed in Figure \ref{fig:trayectories_ej}. Left (right) panel shows trajectories from $P_0$ ($P_1$) and thick solid lines represent empirical means. 
	
	\item Figure \ref{fig:cost} displays the computational cost (in seconds) for different sample sizes $n$ in that example. Each point represents the sum of computation times of 100 experiments for each methodology and sample size with $d=200$. The results have been obtained in a standard PC with processor Intel i7-3820, 3.60 GHz and 32GB RAM. Note that the considered kNN and SVM implementations are computationally efficient (see Section \ref{sec:comp}).
	
	\item Table \ref{table:vs} is a complement for Table 1 by showing the average number of variables (or components).
	
	\item  Tables \ref{t:lda}, \ref{t_knn} and \ref{t_svm} show the classification accuracy (percentage of correct classification) for different groups of models and methods obtained with LDA, kNN and SVM classifiers respectively. 
	Results from the different considered classifiers are quite similar in relative terms. Let us recall that the full results of the $1128$ experiments ($94$ models $\times 4$ samples sizes $\times 3$ classifiers) are available in the supplementary file \textit{outputs}. The methods appear in columns; apart from methods in Table 1 we have included \textit{Base} (except for LDA) and \textit{Oracle} versions of each method. The first is based on  the entire trajectories and \textit{Oracle} only uses the true relevant variables. The simulation outputs  are grouped in different categories (in rows) by model type and sample size $n$. The rows are labelled by the general model type, that is, logistic, Gaussian and mixtures. The logistic models are also divided by the type of processes involved according to the notation given above. RKHS denotes the models that fulfil the hypotheses  of RK-VS (G2, G2b, G4,...,G8) and ``All models'' includes the outputs of all the 94 considered models for each $n$. We have followed the methodology described in the main paper and the outputs are averaged over 200 independent runs. The marked values correspond to the  best performance in each row (excluding \textit{Oracle} which is not feasible in practice). 

\end{itemize}

\

\begin{table}
	\begin{center}
		\caption{\footnotesize Average number of selected variables (or components) with the three considered classifiers. Remember that the original dimension is 100.}\label{table:vs}
		
		\medskip
		
		\footnotesize
		\begin{tabular}{llccccc}
			\toprule
			\small Classifier & \small Sample size & \multicolumn{5}{c}{\small Dimension reduction methods}\\ \cmidrule{3-7} 
			& &  mRMR-RD &  PLS &  MHR &  RK-VS &  RK$_{B}$-VS\\
			\midrule
			\small LDA & $n=30$ & 4.9 & 2.6 & 5.4 & \framebox{2.7} & 3.7 \\
			& $n=50$ & 5.9 & 2.8 & 6.1 & \framebox{2.8} & 4.1  \\
			& $n=100$ & 7.2 & \framebox{3.3} & 7.0 & 3.2 & 4.8  \\
			& $n=200$ & 8.1 & 4.0 & 7.5 & \framebox{3.9} & 5.6 \\
			\midrule
			
			\small kNN & $n=30$ & 7.8 & \framebox{4.3} & 6.2 & 7.6 & 8.1 \\
			& $n=50$ & 8.0 & \framebox{4.8} & 6.2 & 7.3 & 7.9 \\
			& $n=100$ & 8.4 & \framebox{5.5} & 6.2 & 6.7 & 7.6 \\
			& $n=200$ & 8.6 & 6.2 & \framebox{5.9} & 6.3 & 7.2 \\
			\midrule
			
			\small SVM & $n=30$ & 9.3 & 3.3 & 8.0 & 9.3 & 10.0  \\
			& $n=50$ & 9.4 & \framebox{3.8} & 7.9 & 8.7 & 9.6  \\
			& $n=100$ & 9.7 & \framebox{4.6} & 7.9 & 8.0 & 9.2  \\
			& $n=200$ & 9.8 & \framebox{5.6} & 7.5 & 7.6 & 8.9  \\
			\bottomrule
			
		\end{tabular}
	\end{center}
\end{table}

\


\begin{table}
	\begin{center}
		\caption{{\footnotesize Percentage of correct classification with LDA}}{\footnotesize 
			\begin{tabular}{ll|cccccc|c}
				\toprule
				\small Models & \small $n$ & \small mRMR-RD & \small PLS & \small MHR & \small RK-VS & \small RK$_B$-VS & \small Base & \small LDA-Oracle\\
				\midrule
				\small All models & $30$ & 81.04 & \framebox{82.87} & 82.44 & 81.50 & 80.89 & 61.48 & 84.97\\
				& $50$ & 82.37 & \framebox{83.78} & 83.68 & 83.44 & 82.54 & 59.30 & 86.23\\
				& $100$ & 83.79 & 84.70 & 84.97 & \framebox{85.30} & 84.46 & 53.31 & 87.18\\
				& $200$ & 84.88 & 85.46 & 85.90 & \framebox{86.51} & 85.90 & 74.73 & 87.69\\
				\midrule\small Logistic OU & $30$ & 78.70 & \framebox{80.11} & 79.36 & 78.21 & 76.47 & 60.32 & 81.92\\
				& $50$ & 80.12 & \framebox{80.96} & 80.75 & 80.23 & 78.33 & 58.05 & 83.24\\
				& $100$ & 81.70 & 81.90 & \framebox{82.30} & 82.16 & 80.69 & 52.83 & 84.27\\
				& $200$ & 83.05 & 82.74 & 83.65 & \framebox{83.66} & 82.61 & 71.79 & 84.84\\
				\midrule\small Logistic OUt & $30$ & 80.12 & \framebox{81.30} & 80.87 & 79.60 & 78.56 & 61.10 & 83.11\\
				& $50$ & 81.21 & \framebox{82.05} & 81.98 & 81.42 & 80.20 & 58.80 & 84.44\\
				& $100$ & 82.39 & 82.91 & 83.14 & \framebox{83.14} & 82.15 & 53.04 & 85.45\\
				& $200$ & 83.35 & 83.51 & 84.03 & \framebox{84.29} & 83.66 & 73.54 & 85.93\\
				\midrule\small Logistic B & $30$ & 82.79 & \framebox{84.57} & 84.19 & 83.52 & 82.32 & 62.74 & 87.54\\
				& $50$ & 84.18 & 85.55 & 85.59 & \framebox{85.65} & 84.21 & 60.06 & 88.83\\
				& $100$ & 85.74 & 86.60 & 87.16 & \framebox{87.71} & 86.47 & 53.55 & 89.90\\
				& $200$ & 86.88 & 87.50 & 88.33 & \framebox{89.17} & 88.18 & 75.94 & 90.51\\
				\midrule\small Logistic sB & $30$ & 82.95 & \framebox{84.63} & 84.26 & 83.43 & 82.37 & 62.87 & 87.10\\
				& $50$ & 84.18 & \framebox{85.59} & 85.59 & 85.39 & 84.11 & 60.74 & 88.46\\
				& $100$ & 85.51 & 86.60 & 87.02 & \framebox{87.52} & 86.34 & 53.17 & 89.55\\
				& $200$ & 86.71 & 87.38 & 88.20 & \framebox{88.84} & 87.98 & 75.73 & 90.18\\
				\midrule\small Logistic ssB & $30$ & 84.56 & \framebox{85.73} & 85.58 & 84.93 & 84.51 & 63.60 & 86.54\\
				& $50$ & 85.65 & 86.49 & \framebox{86.54} & 86.42 & 85.93 & 60.68 & 87.90\\
				& $100$ & 86.86 & 87.25 & 87.38 & \framebox{87.89} & 87.39 & 53.55 & 88.81\\
				& $200$ & 87.83 & 88.01 & 87.72 & \framebox{88.83} & 88.59 & 75.33 & 89.38\\
				\midrule\small Gaussian & $30$ & 85.28 & 88.63 & 88.70 & 88.30 & \framebox{89.95} & 62.56 & 90.91\\
				& $50$ & 86.72 & 89.45 & 89.38 & 89.81 & \framebox{90.69} & 61.24 & 91.41\\
				& $100$ & 88.21 & 89.91 & 89.86 & 90.81 & \framebox{91.18} & 55.43 & 91.64\\
				& $200$ & 89.00 & 90.38 & 89.96 & 91.13 & \framebox{91.30} & 83.89 & 91.71\\
				\midrule\small Mixture & $30$ & 71.95 & 76.19 & 75.40 & 73.93 & \framebox{76.65} & 55.51 & 79.09\\
				& $50$ & 73.88 & 77.66 & 77.03 & 76.63 & \framebox{78.30} & 55.13 & 80.29\\
				& $100$ & 75.54 & 78.91 & 78.61 & 79.13 & \framebox{79.89} & 52.48 & 81.07\\
				& $200$ & 76.46 & 79.66 & 79.29 & 80.21 & \framebox{80.61} & 70.77 & 81.39\\
				\midrule\small RKHS & $30$ & 85.28 & 88.63 & 88.70 & 88.30 & \framebox{89.95} & 62.56 & 90.91\\
				& $50$ & 86.72 & 89.45 & 89.38 & 89.81 & \framebox{90.69} & 61.24 & 91.41\\
				& $100$ & 88.21 & 89.91 & 89.86 & 90.81 & \framebox{91.18} & 55.43 & 91.64\\
				& $200$ & 89.00 & 90.38 & 89.96 & 91.13 & \framebox{91.30} & 83.89 & 91.71\\
				\bottomrule
				
			\end{tabular}\label{t:lda}}
	\end{center}
\end{table}

\begin{table}
	\begin{center}
		\caption{{\footnotesize Percentage of correct classification with kNN}}{\footnotesize 
			\begin{tabular}{ll|cccccc|c}
				\toprule
				\small Models & \small $n$ & \small mRMR-RD & \small PLS & \small MHR & \small RK-VS & \small RK$_B$-VS & \small Base & \small kNN-Oracle\\
				\midrule
				\small All models & $30$ & 81.88 & 82.45 & \framebox{82.46} & 82.28 & 81.92 & 79.61 & 84.56\\
				& $50$ & 82.95 & 83.49 & 83.43 & \framebox{83.75} & 83.25 & 80.96 & 86.16\\
				& $100$ & 84.31 & 84.77 & 84.73 & \framebox{85.59} & 84.95 & 82.60 & 87.94\\
				& $200$ & 85.38 & 85.79 & 85.91 & \framebox{87.16} & 86.50 & 83.99 & 89.25\\
				\midrule\small Logistic OU & $30$ & 78.71 & \framebox{79.22} & 79.20 & 78.58 & 77.82 & 75.63 & 81.15\\
				& $50$ & 79.64 & \framebox{80.04} & 80.02 & 79.98 & 79.05 & 76.87 & 82.63\\
				& $100$ & 80.96 & 81.13 & 81.26 & \framebox{81.66} & 80.68 & 78.44 & 84.30\\
				& $200$ & 82.10 & 82.07 & 82.56 & \framebox{83.21} & 82.23 & 79.73 & 85.49\\
				\midrule\small Logistic OUt & $30$ & 81.87 & \framebox{82.71} & 82.30 & 81.91 & 81.37 & 79.50 & 84.46\\
				& $50$ & 82.83 & \framebox{83.52} & 83.18 & 83.13 & 82.49 & 80.62 & 85.89\\
				& $100$ & 84.12 & 84.52 & 84.33 & \framebox{84.90} & 84.03 & 82.02 & 87.35\\
				& $200$ & 85.00 & 85.31 & 85.30 & \framebox{86.23} & 85.31 & 83.14 & 88.49\\
				\midrule\small Logistic B & $30$ & 83.29 & \framebox{84.01} & 83.94 & 83.94 & 83.04 & 81.10 & 86.61\\
				& $50$ & 84.38 & 85.08 & 84.90 & \framebox{85.47} & 84.55 & 82.35 & 88.24\\
				& $100$ & 85.68 & 86.30 & 86.31 & \framebox{87.40} & 86.41 & 83.92 & 90.19\\
				& $200$ & 86.78 & 87.39 & 87.63 & \framebox{89.27} & 88.25 & 85.35 & 91.66\\
				\midrule\small Logistic sB & $30$ & 84.00 & 84.48 & \framebox{84.55} & 84.40 & 83.66 & 81.90 & 86.59\\
				& $50$ & 84.87 & 85.36 & 85.31 & \framebox{85.65} & 84.93 & 83.02 & 88.24\\
				& $100$ & 86.09 & 86.61 & 86.62 & \framebox{87.51} & 86.62 & 84.44 & 90.11\\
				& $200$ & 87.07 & 87.58 & 87.84 & \framebox{89.17} & 88.35 & 85.73 & 91.59\\
				\midrule\small Logistic ssB & $30$ & 85.92 & 85.97 & 86.35 & \framebox{86.39} & 86.09 & 84.47 & 88.01\\
				& $50$ & 86.86 & 86.78 & 87.11 & \framebox{87.49} & 87.10 & 85.41 & 89.44\\
				& $100$ & 87.93 & 87.86 & 88.05 & \framebox{88.89} & 88.55 & 86.71 & 91.04\\
				& $200$ & 88.89 & 88.81 & 88.75 & \framebox{90.24} & 89.88 & 87.91 & 92.34\\
				\midrule\small Gaussian & $30$ & 83.96 & 85.35 & 85.79 & 86.16 & \framebox{87.13} & 83.20 & 87.46\\
				& $50$ & 84.80 & 86.61 & 86.68 & 87.62 & \framebox{88.20} & 84.99 & 88.55\\
				& $100$ & 85.69 & 87.85 & 87.58 & 88.91 & \framebox{89.19} & 86.61 & 89.56\\
				& $200$ & 86.30 & 88.74 & 88.19 & 89.68 & \framebox{89.84} & 87.94 & 90.11\\
				\midrule\small Mixture & $30$ & 74.20 & 74.40 & 74.40 & 74.42 & \framebox{75.92} & 71.05 & 76.83\\
				& $50$ & 76.59 & 76.92 & 76.70 & 77.45 & \framebox{78.43} & 73.92 & 79.58\\
				& $100$ & 79.46 & 79.68 & 79.20 & 80.76 & \framebox{81.36} & 77.32 & 82.70\\
				& $200$ & 81.48 & 81.51 & 81.42 & 83.21 & \framebox{83.61} & 79.98 & 84.74\\
				\midrule\small RKHS & $30$ & 83.96 & 85.35 & 85.79 & 86.16 & \framebox{87.13} & 83.20 & 87.46\\
				& $50$ & 84.80 & 86.61 & 86.68 & 87.62 & \framebox{88.20} & 84.99 & 88.55\\
				& $100$ & 85.69 & 87.85 & 87.58 & 88.91 & \framebox{89.19} & 86.61 & 89.56\\
				& $200$ & 86.30 & 88.74 & 88.19 & 89.68 & \framebox{89.84} & 87.94 & 90.11\\
				\bottomrule
				
			\end{tabular}\label{t_knn}}
	\end{center}
\end{table}

\begin{table}
	\begin{center}
		\caption{{\footnotesize Percentage of correct classification with SVM}}
		{\footnotesize \begin{tabular}{ll|cccccc|c}
				\toprule
				\small Models & \small $n$ & \small mRMR-RD & \small PLS & \small MHR & \small RK-VS & \small RK$_B$-VS & \small Base & \small SVM-Oracle\\
				\midrule
				\small All models & $30$ & 83.22 & 84.12 & \framebox{84.62} & 84.28 & 84.12 & 83.86 & 87.53\\
				& $50$ & 84.21 & 85.04 & 85.44 & \framebox{85.60} & 85.20 & 85.01 & 88.21\\
				& $100$ & 85.27 & 86.03 & 86.29 & \framebox{86.96} & 86.48 & 86.20 & 88.75\\
				& $200$ & 86.10 & 86.79 & 86.86 & \framebox{87.90} & 87.50 & 87.07 & 89.03\\
				\midrule\small Logistic OU & $30$ & 79.98 & 80.79 & \framebox{80.81} & 80.19 & 79.65 & 80.18 & 83.93\\
				& $50$ & 81.13 & 81.64 & \framebox{81.69} & 81.66 & 80.95 & 81.36 & 84.62\\
				& $100$ & 82.39 & 82.51 & 82.59 & \framebox{83.15} & 82.44 & 82.50 & 85.17\\
				& $200$ & 83.51 & 83.30 & 83.50 & \framebox{84.32} & 83.74 & 83.42 & 85.49\\
				\midrule\small Logistic OUt & $30$ & 83.38 & 83.84 & \framebox{84.33} & 83.70 & 83.28 & 83.77 & 87.24\\
				& $50$ & 84.37 & 84.69 & \framebox{85.14} & 85.00 & 84.39 & 84.82 & 87.88\\
				& $100$ & 85.43 & 85.67 & 86.07 & \framebox{86.34} & 85.75 & 85.94 & 88.37\\
				& $200$ & 86.15 & 86.34 & 86.71 & \framebox{87.26} & 86.74 & 86.71 & 88.64\\
				\midrule\small Logistic B & $30$ & 85.24 & 85.81 & \framebox{87.01} & 86.56 & 85.97 & 86.01 & 90.58\\
				& $50$ & 86.23 & 86.83 & 87.92 & \framebox{88.11} & 87.20 & 87.17 & 91.23\\
				& $100$ & 87.35 & 87.92 & 88.99 & \framebox{89.58} & 88.69 & 88.50 & 91.80\\
				& $200$ & 88.16 & 88.85 & 89.85 & \framebox{90.71} & 89.95 & 89.50 & 92.09\\
				\midrule\small Logistic sB & $30$ & 85.55 & 85.98 & \framebox{87.06} & 86.68 & 86.22 & 86.22 & 90.22\\
				& $50$ & 86.33 & 86.96 & \framebox{87.92} & 87.86 & 87.32 & 87.32 & 90.96\\
				& $100$ & 87.13 & 88.01 & 88.88 & \framebox{89.41} & 88.69 & 88.51 & 91.53\\
				& $200$ & 88.04 & 88.84 & 89.55 & \framebox{90.41} & 89.80 & 89.40 & 91.81\\
				\midrule\small Logistic ssB & $30$ & 87.16 & 87.31 & 87.69 & \framebox{88.26} & 88.25 & 87.65 & 90.08\\
				& $50$ & 87.93 & 88.02 & 88.28 & \framebox{89.07} & 88.90 & 88.47 & 90.57\\
				& $100$ & 88.82 & 88.96 & 88.55 & \framebox{89.91} & 89.77 & 89.37 & 91.00\\
				& $200$ & 89.47 & 89.73 & 88.54 & \framebox{90.60} & 90.57 & 90.16 & 91.25\\
				\midrule\small Gaussian & $30$ & 86.42 & 88.72 & 88.97 & 89.00 & \framebox{89.99} & 87.29 & 90.54\\
				& $50$ & 87.33 & 89.44 & 89.27 & 89.94 & \framebox{90.49} & 88.81 & 91.02\\
				& $100$ & 88.48 & 90.03 & 89.60 & 90.63 & \framebox{90.93} & 89.88 & 91.38\\
				& $200$ & 88.98 & 90.41 & 89.51 & 91.03 & \framebox{91.21} & 90.48 & 91.45\\
				\midrule\small Mixture & $30$ & 73.01 & 76.52 & 76.12 & 75.53 & \framebox{76.93} & 74.88 & 78.71\\
				& $50$ & 74.39 & 77.90 & 77.42 & 77.50 & \framebox{78.35} & 76.51 & 79.89\\
				& $100$ & 75.55 & 79.27 & 78.65 & 79.41 & \framebox{79.72} & 78.20 & 80.76\\
				& $200$ & 76.35 & 80.10 & 79.06 & 80.26 & \framebox{80.50} & 79.21 & 81.16\\
				\midrule\small RKHS & $30$ & 86.42 & 88.72 & 88.97 & 89.00 & \framebox{89.99} & 87.29 & 90.54\\
				& $50$ & 87.33 & 89.44 & 89.27 & 89.94 & \framebox{90.49} & 88.81 & 91.02\\
				& $100$ & 88.48 & 90.03 & 89.60 & 90.63 & \framebox{90.93} & 89.88 & 91.38\\
				& $200$ & 88.98 & 90.41 & 89.51 & 91.03 & \framebox{91.21} & 90.48 & 91.45\\
				\bottomrule
				
			\end{tabular}\label{t_svm}}
	\end{center}
\end{table}

\end{document}